\documentclass[12pt]{iopart}
\usepackage{iopams}  
\usepackage{graphicx}
\usepackage{subfig}
\usepackage{tikz}

\usepackage[symbol]{footmisc}

\input{epsf}
\newcommand{\bea}{\begin{eqnarray}}
\newcommand{\eea}{\end{eqnarray}}
\newcommand{\be}{\begin{equation}}
\newcommand{\ee}{\end{equation}}

\begin{document}

\title[$\zeta$-function for a model with spectral dependent boundary conditions]{$\zeta$-function for a model with spectral dependent boundary conditions}%

\author{\underline{Horacio Falomir}}%
\address{Instituto de Física de La Plata (IFLP), CONICET and Universidad Nacional de La Plata (UNLP), CC 67, 1900 La Plata, Argentina}%
\ead{falomir@fisica.unlp.edu.ar\hfill\break
(Corresponding author)}%

\author{Marcelo Loewe}
\address{Facultad de Ingenier\'ia, Arquitectura y Diseño, Universidad San Sebasti\'an, Santiago, Chile.}
\address{Centre for Theoretical and Mathematical Physics, and Department of Physics, University of Cape Town, Rondebosch 7700, South Africa.}
\ead{marcelo.loewe@uss.cl}%

\author{Enrique Mu\~{n}oz}%
\address{Facultad de F\'isica, Pontificia Universidad Cat\'olica de Chile, Vicu\~na Mackenna 4860, Santiago, Chile.}%
\address{Center for Nanotechnology and Advanced Materials CIEN-UC, Vicuña Mackenna 4860, Santiago, Chile}
\ead{ejmunozt@uc.cl}%

\author{Juan Crist\'obal Rojas}%
\address{Departamento de F\'isica, Universidad Cat\'olica del Norte, Angamos 610, Antofagasta, Chile}
\ead{jurojas@ucn.cl}

\vspace{2pc}

\date{\today}%
\begin{abstract}
We explore the meromorphic structure of the $\zeta$-function associated with the boundary eigenvalue problem of a modified Sturm-Liouville operator subject to spectral-dependent boundary conditions at one end of a segment of length $l$. We find that it presents isolated simple poles that follow the general rule valid for second-order differential operator subject to standard local boundary conditions. We employ our results to evaluate the determinant of the operator and the Casimir energy of the system it describes, and study its dependence on $l$ for both the massive and the massless cases. 
\end{abstract}
\maketitle
\section{Introduction}
The field theoretical description of physical systems with spatial boundaries or discontinuities in the local parameters of the Lagrangian density leads to the consideration of differential operators with non-trivial boundary conditions involving the dynamics of the corresponding observables. Moreover, the secular eigenvalue problem can incorporate an explicit spectral dependence in these boundary conditions. An example of this situation was introduced in Ref.~\cite{Fosco1,Wilson}, where a simple model for a one-dimensional waveguide of total length $l$, terminated on a superconducting quantum interference device (SQUID), is represented by a scalar field $\varphi(t,z)$ (corresponding to the magnetic flux at each point $z$ along the line) subject to a boundary condition at the end of this segment which involves the second time (tangent) derivative of the field. Ref.~\cite{Fosco1} deals with the Casimir effect \cite{Bordag,Asorey,Dodonov} for these kind of dynamical boundary conditions.  

This model has recently inspired the consideration of a somewhat more general dynamical boundary conditions in \cite{JA_W-1}, which also involves the first time-derivative of the scalar field at $z=l$, describing the coupled dynamics of the bulk quantum field with a boundary degree of freedom. The Fock states for this system are explicitly constructed and the renormalized local vacuum energy (mean value of the Hamiltonian density) is defined through a Hadamard subtraction, a result which is later employed to (numerically) estimate, by integration on the space slice,  the Casimir energy of the system. In \cite{JA_W-2} this construction is extended to the positive temperature case.    

 Particular motivations of these articles have been, for example, the interest in studying the Casimir effect \cite{Bordag,Asorey,Dodonov} in conditions related to realistic experimental settings as in \cite{Fosco1,Wilson}, or to contribute to the notion of boundary degrees of freedom \cite{Barbero1,Barbero2,Dappiaggi} in connection with holography in high-energy physics or condensed matter physics.

 When looking for stationary solutions, these dynamical equations reduce to a \emph{boundary eigenvalue problem}, a second-order differential operator subject to boundary conditions at $z=l$ which depend on the eigenvalue parameter. This kind of modified Sturm-Liouville problem has been previously considered both in Mathematics and Physics \cite{Walter,Fulton,Mennicken,Barbero1,Zhan,Zhang} in relation with several applications of interest (See \cite{JA_W-2,JA_W-3} for a thorough description of the references on the subject).  In higher dimensions, boundary conditions depending on the Laplacian in the tangential variables of the trace of a scalar field on a hyperplane have been considered in \cite{Relative-Zeta}.

 It is the aim of the present article to explore the meromorphic structure of the $\zeta$-function associated to this kind of boundary eigenvalue problem and employ it to evaluate some physically interesting objects as the functional determinant and the (global) Casimir energy of these systems. Indeed, spectral functions — such as the $\zeta$-function and the heat-trace — encode relevant information about the quantum properties of the related field theory \cite{Seeley1,Seeley2,Gilkey,Ray-Singer,Dowker,Hawking,Klaus}, such as the one-loop contribution to the effective action or the dependence of the vacuum energy on external conditions. 

\bigskip

 In the next Section 
 we introduce the Hilbert space on which is defined the theory. In Section \ref{spectrum} we analyze the spectrum of the operator for the massive case and determine the singularities of the meromorphic prolongation of the $\zeta$-function based on the large eigenvalue behavior. We evaluate the functional determinant as well as the dominant contributions to the Casimir energy and specify the renormalization needed. 
 
 In the rest of the paper we employ an alternative representation of the $\zeta$-function as an integration on a path on the complex plane encircling the spectrum, and employ it to get the poles and residues as well as the regular parts expressed in a form adequate for numerical evaluation. We give the determinant and the Casimir energy both for the massive and massless case, and analyze its behavior for large and small $l$ as a function of the parameters in the boundary condition. In order to simplify the exposition, we add some appendices with auxiliary calculations. We also include in the last appendix some considerations on a heuristic toy model to further motivate the study of this kind of dynamical boundary conditions. 

 In Section \ref{Conclusions} we summarize our conclusions.

\section{The dynamical equation}\label{dynamical_EoM}

In reference \cite{JA_W-1}, the quantization of a mixed bulk-boundary system describing the coupled dynamics of a \emph{bulk} scalar quantum field confined to a segment $z\in [0,l]$ and a \emph{boundary observable} corresponding to its value at $z=l$ was reduced to the consideration of a self-adjoint operator $A$ defined on an enlarged Hilbert space \cite{Walter,Fulton,Zhang}
\begin{equation}\label{1}
    \mathcal{H}:=L^2\left( [0,l] \right) \oplus \mathbb{C},
\end{equation}
with elements of the form
\begin{equation}\label{2}
    \varphi(z)=\left(
                 \begin{array}{c}
                   \varphi_1(z) \\
                   \varphi_2 \\
                 \end{array}
               \right)
\end{equation}
and scalar product defined by
\begin{equation}\label{3}
    \left(\varphi,\chi \right)_{\mathcal{H}}:= \left(\varphi_1,\chi_1 \right)_{L^2\left( [0,l] \right)}+\frac{1}{\rho}\,{\varphi_2}^* \chi_2,
\end{equation}
where $\rho>0$. The operator $A$ is defined on the domain
\begin{equation}\label{4}
\begin{array}{c}\displaystyle
  \mathcal{D}(A):=\left\{ \varphi(z)\in \mathcal{H} : \varphi_1(z), \varphi_1'(z) \in\mathcal{AC}[0,l], \varphi_1''(z)\in L^2\left( [0,l] \right),\right.
  \\ \\ \displaystyle
  \left. \cos\alpha \, \varphi_1(0) + \theta \sin\alpha \, \varphi_1'(0) =0, \varphi_2=\beta_1' \varphi_1(l)-\beta_2' \varphi_1'(l)  \right\}
\end{array}
\end{equation}
with $\alpha\in[0,\pi)$ and $\theta$ a constant with length units, where it acts as
\begin{equation}\label{5}
    A\varphi(z):=\left(
                   \begin{array}{c}
                     \left[-{\partial_z}^2+m^2+V(z) \right] \varphi_1(z) \\
                     -\left[ \beta_1 \varphi_1(l)-\beta_2 \varphi_1'(l) \right] \\
                   \end{array}
                 \right)
\end{equation}
where $V(z)$ is a bounded function. The dynamical equation of the coupled system reads as
\begin{equation}\label{6}
   \left( {\partial_t}^2 + A  \right) \varphi(z)=0,
\end{equation}
which implies that the field satisfy the differential equation
\begin{equation}\label{7}
    \left[ {\partial_t}^2 -{\partial_z}^2+m^2+V(z)   \right] \varphi(z)=0
\end{equation}
and dynamical boundary conditions
\begin{equation}\label{8}
    \begin{array}{c} \displaystyle
    \cos\alpha \, \varphi_1(t,0) +\theta  \sin\alpha \, \partial_z\varphi_1(t,0) =0,
      \\ \\ \displaystyle
      {\partial_t}^2 \left[\beta_1' \varphi_1(t,l)-\beta_2' \partial_z\varphi_1(t,l)\right]=
      \left[ \beta_1 \varphi_1(t,l)-\beta_2 \partial_z\varphi_1(t,l) \right].
    \end{array}
\end{equation}
Note that the parameters $\beta$ in the second line of Eq.~(\ref{8}) can be simultaneously scaled without altering their significance.

Therefore, for the stationary states we write $\varphi(t,z)$ as $e^{-i \omega t} \varphi(z), \varphi\in\mathcal{D}(A)$, and get
\begin{equation}\label{9}
    \begin{array}{c} \displaystyle
    A \varphi(z)=\omega^2 \varphi(z)
      \\ \\ \displaystyle
      \cos\alpha \, \varphi_1(0) +\theta  \sin\alpha \, \partial_z\varphi_1(0) =0,
      \\ \\ \displaystyle
      \left( \beta_2+\omega^2\beta_2'\right) \varphi_1'(l)=
      \left( \beta_1 +\omega^2 \beta_1' \right) \varphi_1(l),
    \end{array}
\end{equation}
\emph{i.e.}, a \emph{boundary eigenvalue problem} with a boundary condition dependent on the spectral parameter.

It has been proved \cite{Walter,Fulton,JA_W-1,Zhang} that $A$ so defined is self-adjoint if\footnote{Notice that, if $l\sim  {\rm mass}^{-1}$, $\beta_1 \sim {\rm mass}^2$, $\beta_2 \sim {\rm mass}^1$, $\beta_1' \sim {\rm mass}^{0}$, $\beta_2' \sim {\rm mass}^{-1}$ and $\rho \sim {\rm mass}$.} 
\begin{equation}
\rho= \beta_1' \beta_2-\beta_1\beta_2'>0,    
\end{equation}
 has a discrete spectrum of multiplicity one eigenvalues that accumulate at $\infty$, and is positive if $m^2+V(z) >0$ for $z\in[0,l]$, $\alpha=0$ or $\alpha \in [\frac{\pi}{2},\pi)$, and
\begin{equation}\label{10}
    \begin{array}{c} \displaystyle
    \beta_1\geq 0,\quad \beta_1',\beta_2<0,\quad {\rm for}\quad \beta_2'>0,
        \\ \\ \displaystyle
    \beta_1\leq 0,\quad \beta_1',\beta_2>0,\quad {\rm for}\quad \beta_2'<0,
    \end{array}
\end{equation}
where the second line is obtained from the first one through a simultaneous change of sign of these four parameters.  Our study will also be applicable to the case $\beta'_2 =0$ \cite{Fosco1}. So, in the following we will take $\beta_2'\geq 0$.

Under these conditions, the smallest eigenvalue of $A$ is positive, ${\omega_1}^2>0$. Moreover, for $\beta_1\neq 0$, it is $\beta_1' \beta_1<0$ and $\frac{\beta_2'}{\beta_1'}<0$. Then, $\rho>0$ implies that
\begin{equation}\label{10-1}
    \frac{\beta_2}{\beta_1}<\frac{\beta_2'}{\beta_1'}<0. \quad
\end{equation}

Note that since $A$ is unbounded, for large eigenvalues $\omega$  the last line of Eq.~(\ref{9}) effectively enforces standard local boundary conditions at $z=l$ determined by $\beta'_1$ and $\beta'_2$: generalized Robin boundary conditions for $\beta'_2\neq 0$ and Dirichlet boundary conditions for $\beta'_2=0$. This suggests that the meromorphic structure of $\zeta_A(s)$, the $\zeta$-function associated with this operator, governed by the large eigenvalue behavior, is similar to that of the Sturm-Lioville operator under these local boundary conditions, which is entirely determined by the dimension of the manifold and the order of the differential operator \cite{Seeley1,Seeley2,Gilkey}.

Since our primary interest lies in understanding the effects of imposing these generalized boundary conditions at $z=l$, for the sake of clarity in the exposition, in the following we take $V(z)\equiv 0$ and $\alpha=0$, thereby imposing Dirichlet boundary conditions at the origin, $\varphi_1(0)=0$. 

In the following sections, we determine the spectrum of the boundary eigenvalue problem in 
Eq.~(\ref{9}), construct $\zeta_A(s)$ for both the massive and massless cases, and demonstrate that it is holomorphic in a neighborhood of the origin. This property allows for the standard definition of the functional determinant of $A$ as $e^{-\zeta_A'(0)}$.

We also show that, in both cases, $\zeta_A(s)$ exhibits a simple pole at $s = -1/2$, which implies that using this regularization to define the Casimir energy of the model requires renormalizing a constant term (a redefinition of the reference energy level) and, in the massive case, a term linear in the segment length $l$ (i.e., the subtraction of a singular energy density).

Finally, we present integral representations of the finite contributions to the Casimir energies and analyze their behavior as functions of $l$.

\section{The spectrum}\label{spectrum}

The general solution of
\begin{equation}\label{11}
    \left( -{\partial_z}^2+m^2\right)\varphi_1(z)=\omega^2 \varphi_1(z), \quad {\rm with}\quad \varphi_1(0)=0
\end{equation}
is given by
\begin{equation}\label{12}
   \varphi_1(z) \sim \sin \left( z \sqrt{\omega^2 -m^2}\right),
\end{equation}
where $\omega^2>0$ since we are considering a positive definite operator. Then, the boundary condition at $z=l$ gives
\bea\label{13}
    &&\left( \beta_2+\omega^2\beta_2'\right)  \sqrt{\omega^2 -m^2} \cos \left( l \sqrt{\omega^2 -m^2}\right)\nonumber\\
    &&=
      \left( \beta_1 +\omega^2 \beta_1' \right) \sin \left( l \sqrt{\omega^2 -m^2}\right).
\eea

\medskip

For $0< \omega^2 < m^2$, the boundary condition at $z=l$ gives
\bea\label{12-1}
    &&\left( \beta_2+\omega^2\beta_2'\right) i \sqrt{m^2 - \omega^2 } \cosh \left( l \sqrt{m^2-\omega^2 }\right)\nonumber\\
    &&= \left( \beta_1 +\omega^2 \beta_1' \right) i \sinh \left( l \sqrt{m^2-\omega^2 }\right),
\eea
or
\begin{equation}\label{12-2}
    \frac{\tanh \left( l \sqrt{m^2-\omega^2 }\right)}{l \sqrt{m^2-\omega^2 }} = \frac{(\beta_2+\omega^2\beta_2')}{l(\beta_1 +\omega^2 \beta_1')},
\end{equation}
which has no solutions since the left hand side is positive for $0< \omega^2 < m^2$, while the right hand side is a decreasing function of $\omega^2$ that takes the value $\frac{\beta_2}{l \beta_1}<0$ for $\omega^2=0$. Indeed,
\begin{equation}\label{12-3}
    \frac{\partial}{\partial \omega^2} \left(\frac{(\beta_2+\omega^2\beta_2')}{l(\beta_1 +\omega^2 \beta_1')}\right)=
    \frac{-\rho}{l(\beta_1 +\omega^2 \beta_1')^2}<0.
\end{equation}

For $\omega^2 \geq m^2$, defining $x:=l \sqrt{\omega^2 -m^2}$, the spectrum is given by the zeroes of
\begin{equation}\label{14}
    f(x):=x \left(a+b \, x^2 \right) \cos x - \left(c+d \, x^2 \right) \sin x
\end{equation}
where we have defined the dimensionless parameters
\begin{equation}\label{abcd}
    a=l(\beta_2+m^2\beta_2'),\  b=\beta_2'/l, \ c=l^2(\beta_1 +m^2 \beta_1')\  {\rm and}\  d=\beta_1',
\end{equation}
with $l \rho=(a d-b c)>0$.

Note that $x=0$  ($\omega=m$) does not correspond to an eigenvalue even though $f(0)=0$, since $\varphi_1(z)$ in Eq.~(\ref{12}) is identically vanishing in this case. Therefore, the first eigenvalue of $A$ is ${\omega_1}^2>m^2$.

In the following in this Section we will assume $\beta_2'\neq 0$ ($b\neq0$). The case $\beta_2' =0$ will be considered later.

\medskip

We can approximate the large eigenvalues of $A$ taking into account that $f(x)=0$ implies that
\begin{equation}\label{15}
    \frac{ \tan x  }{x}= \frac{a+b \, x^2 }{c+d \, x^2 } \approx \left\{
    \begin{array}{l}
      \frac{a}{c}  +O(x^2), \quad {\rm for} \quad x<<1, \, c \neq 0,
      \\ \\
       \frac{b}{d}+ O(x^{-2})<0, \quad {\rm for }\quad x \rightarrow \infty,\, d \neq 0,
    \end{array}
    \right.
\end{equation}
where the rational function in the right hand side is a decreasing function,
\begin{equation}\label{16}
    \frac{\partial}{\partial x}\left( \frac{ \left(a+b \, x^2 \right)}{ \left(c+d \, x^2 \right) } \right)
    = \frac{-2 l \rho x}{ \left(c+d \, x^2 \right)^2}<0.
\end{equation}

\begin{figure} \label{ceros.fig}
\includegraphics[width=\linewidth]{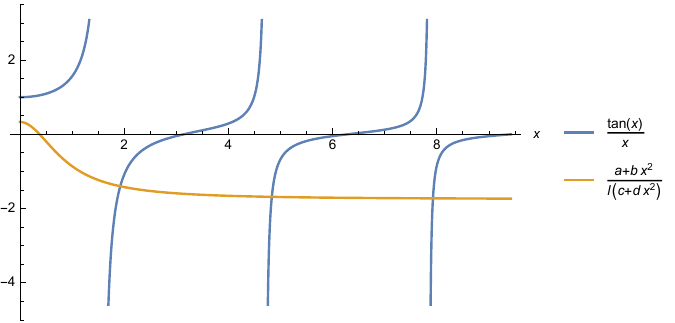}
\caption{The spectrum ($l=1,a=-1,b=7,c=-3,d=-4$).}
\end{figure}

Since the asymptotes of the left hand side of Eq.~(\ref{15}) are located at $(n-1/2)\pi, n=1,2,3,\cdots$, we posit an asymptotic expansion for the solutions of that equation of the form
\begin{equation}\label{17}
    x_n\asymp (n-1/2)\pi + \sum_{k\geq 1}^K \frac{y_k}{n^k} + O\left(n^{-K-1}\right),\quad n\in \mathbb{N},
\end{equation}
and iteratively solve for the coefficients to get
\begin{equation}\label{18}
    \begin{array}{l}\displaystyle
      y_1=-\frac{d }{\pi  b},
       \\ \\ \displaystyle
      y_2=- \frac{d }{2 \pi  b},
      \\ \\ \displaystyle
       y_3= \frac{ \left[12 a b d-3 b^2 \left(4 c+\pi ^2
   d\right)\right]}{12 \pi ^3 b^3}-\frac{d^2 }{\pi ^3
   b^2}+\frac{d^3 }{3 \pi ^3 b^3},
   \\ \\ \displaystyle
   y_4= \frac{\left[12 a b d-b^2 \left(12 c+\pi ^2
   d\right)\right]}{8 \pi ^3 b^3}-\frac{3 d^2 }{2 \pi^3 b^2}+\frac{d^3 }{2 \pi ^3 b^3},
   \\ \\
   \cdots
    \end{array}
\end{equation}

Note also that, for $\frac{a}{c}\leq 1$, there are no solutions of Eq.~(\ref{15}) with $x<\frac{\pi}{2}$. If this is not the case, there is an additional solution $x_0$ which should be numerically evaluated and its contribution explicitly added.

\medskip

This asymptotic expansion of the eigenvalues of $A$ is sufficient to determine the nearest singularities of the analytic extension of the associated $\zeta$-function \cite{Gilkey}, defined as\footnote{Ref.\ \cite{Relative-Zeta} deals with operators that have a non-empty essential spectrum, overcoming this difficulty by defining a "relative" $\zeta$-function.  Since our operator has a discrete, non-degenerate spectrum with an accumulation point at infinity, the standard definition of the $\zeta$-function is suitable.} 
\begin{equation}\label{19}
    \zeta_A(s):= \sum_{n=1}^{\infty} \left(\frac{\omega_n}{\mu}\right)^{-2s} 
\end{equation}
for $\Re s$ large enough, where $\mu$ is an \emph{arbitrarily chosen mass scale}. In terms of $x_n= l \sqrt{{\omega_n}^2 -m^2}$,
\begin{equation}\label{20}
    \zeta_A(s)= (\mu l)^{2s}\sum_{n=1}^{\infty} \left({l^2{\omega_n}^2}\right)^{-s} =
    (\mu l)^{2s}\sum_{n=1}^{\infty} \left({x_n}^2+M^2\right)^{-s},
\end{equation}
where we have defined $M:=l m$.

Notice that Eq.~(\ref{17}) implies that this series is absolutely and uniformly convergent (and therefore analytic) for $\Re s >\frac{1}{2}$. From Eqs.~(\ref{17}) and (\ref{18}), we have
\bea\label{21}
      &&\left({x_n}^2+M^2\right)^{-s} = \left\{\left[\left(n-\frac{1}{2}\right)\pi
      +\frac{{y_1}}{n} +\frac{{y_2}}{n^2}+
   \frac{\Delta}{n^3}\right]^2+M^2\right\}^{-s}\nonumber\\
      &&= \pi^{-2s} \Bigg\{n^{-2 s}+s n^{-2s-1}+
      \frac{s  \left[8 d -4 b M^2+\pi ^2 b (2s+1)\right]}{4 \pi^2 b}\, n^{-2 s-2}\nonumber\\
      &&+O\left(n^{-2 s-3}\right)\Bigg\} ,
\eea
where $\Delta = O(1)$. Then,
\begin{equation}\label{22}
    \begin{array}{c}\displaystyle
      \zeta_A(s)=
    \left(\frac{\mu l}{\pi}\right)^{2s}\left\{\zeta(2 s)+s\, \zeta(2 s+1)+ \phantom{\frac{A}{B}}\right.
    \\ \\ \displaystyle
      \left. +\frac{s \left[8 d -4 b M^2+\pi ^2 b (2 s+1)\right]}{4 \pi ^2 b} \, \zeta(2s+2)+ \triangle\zeta(s)\right\},
    \end{array}
\end{equation}
where $\zeta(s)$ is the Riemann $\zeta$-function (which is a meromorphic function on the whole complex plane, which is holomorphic everywhere except for an isolated simple pole at $s = 1$ with residue $1$ \cite{AS}), and $\triangle\zeta(s)$ is analytic in the half-plane $\Re s >-1$,
\begin{equation}\label{23}
    \begin{array}{c}\displaystyle
      \triangle\zeta(s)=(\mu l)^{2s}\sum_{n=1}^{\infty} \left[\left({x_n}^2+M^2\right)^{-s} - \right.
      \\ \\ \displaystyle
      \left.
      \pi^{-2s}n^{-2 s} \left\{ 1+\frac{s}{n} +
      \frac{s  \left[8 d -4 b M^2+\pi ^2 b (2s+1)\right]}{4 \pi^2 b n^2} \right\}
      \right]
    \end{array}
\end{equation}

Notice that, in particular, $\triangle\zeta(s)$ is analytic at the origin and
\begin{equation}\label{24}
        \begin{array}{c}\displaystyle
        \triangle\zeta(0)=0,
        \\ \\ \displaystyle
      \triangle\zeta'(0)= - \sum_{n=1}^{\infty} \left[
      \log \left(\frac{{x_n}^2+M^2}{\pi^2 n^2}\right)+\frac{1}{n}+\frac{-4 b M^2+\pi ^2 b+8 d }{4 \pi ^2 b n^2} \right],
    \end{array}
\end{equation}
which is an absolutely convergent series (which could be evaluated numerically, for example).

\medskip

The knowledge of $\zeta_A(s)$ enables the evaluation of several relevant magnitudes.
In this context, the determinant of the operator $A$ is defined as \cite{Ray-Singer,Dowker,Hawking}
\begin{equation}\label{25}
    \begin{array}{c} \displaystyle
     \log {\rm Det}(A):={-{\zeta_A}'(0)}= \log\left(\frac{\mu l}{\pi}\right) +
      \\ \\ \displaystyle
      +\frac{l^2 m^2}{6}-\frac{d }{3 b}-\frac{\pi^2}{24}-\gamma +\log (2 \pi )-\triangle\zeta'(0).
    \end{array}
\end{equation}

\medskip

The vacuum energy of the quantum system, $\sum_n \frac{\hbar\omega_n}{2}$, is a formally divergent quantity that requires a precise definition through regularization. In the context of this analytical regularization scheme, and relative to the chosen reference mass scale $\mu$, the Casimir energy is defined as the analytic continuation
\begin{equation}\label{26}
    E_{Cas}^{(0)}(l) :=
    \left.\frac{\hbar \mu}{2} \sum_n \left(\frac{\omega_n}{\mu}\right)^{-2s}\right|_{s\rightarrow -\frac{1}{2}}
    =\left.\frac{\hbar \mu}{2} \, \zeta_A(s)\right|_{s\rightarrow -\frac{1}{2}}.
\end{equation}
In our case, from Eq.~(\ref{22}), we see that the $\zeta$-function has a simple pole at  $s=-1/2$,
\begin{equation}\label{27}
    \begin{array}{c} \displaystyle
      \zeta_A(s)=
      -\frac{  \beta_1'}{2 \pi \mu \beta_2'   \left(s+\frac{1}{2}\right)}+
       \frac{  m^2 l }{4 \pi \mu    \left(s+\frac{1}{2}\right)}
      \\ \\ \displaystyle
      +\frac{\pi }{24 l \mu }
      + \frac{\beta_1' }{\pi \mu \beta_2'  }\left[1-\gamma -\log\left(\frac{l \mu}{\pi} \right)\right]
      \\ \\ \displaystyle
      +\frac{l m^2 \left[\log \left(\frac{l \mu}{\pi }\right)+\gamma -1\right]}{2 \pi\mu } +\triangle\zeta(-1/2) + O\left(s+\frac{1}{2}\right).
    \end{array}
\end{equation}

Therefore, it is also necessary to {\it renormalize} the vacuum energy by introducing counterterms to cancel the singular terms in Eq.~(\ref{27}). Subtracting the singular term that is independent of $l$ has no physical consequences, as it merely redefines the reference level of the energy. However, the residue also contains a linear term in $l$ corresponding to a divergent energy density. Subtracting this term leaves a (finite) linear contribution with an \emph{undetermined} coefficient. Then, the Casimir energy should be defined as
\begin{equation}\label{28}
    \begin{array}{c} \displaystyle
      E_{Cas}(l):= \mathcal{E}_0 + \mathcal{E}_1 l
      +\frac{\pi\hbar }{48 l  }
      + \frac{ \hbar \beta_1' }{2\pi  \beta_2'  }\left[1-\gamma -\log\left(\frac{l \mu}{\pi} \right)\right]+
      \\ \\ \displaystyle
      +\frac{l m^2 \hbar\left[\log \left(\frac{l \mu}{\pi }\right)+\gamma -1\right]}{4 \pi } +\frac{\hbar \mu}{2}\, \triangle\zeta(-1/2),
    \end{array}
\end{equation}
where $\mathcal{E}_0$ is a finite constant (which could be set to 0), $\mathcal{E}_1$ is a \emph{phenomenological} parameter (which should be determined \emph{experimentally}) and
\begin{equation}\label{DELTA_E}
    \begin{array}{c} \displaystyle
        \frac{\hbar \mu}{2}\,\triangle\zeta(-1/2)= \\ \\
         \displaystyle
        =\frac{\hbar}{2 l}\sum_{n=1}^{\infty} \left[\left({x_n}^2+M^2\right)^{1/2} -
  \pi n  \left\{ 1-\frac{1}{2n} -
      \frac{\left[2 d l- b m^2 l^2\right]}{2 \pi^2 b n^2} \right\}
      \right]
    \end{array}
\end{equation}
is an absolutely convergent series since, from Eqs.~(\ref{17}) and (\ref{18}), the general term behaves as $n^{-2}$ for large $n$.

In Section \ref{ECasimir}, we show that $E_{Cas}(l)$ exhibits a linear term that dominates as $l \rightarrow \infty$ (unless it vanishes). Thus, $\mathcal{E}_1$ represents an energy density that drives the expansion or contraction of the manifold (segment) for large values of $l$. In this sense, it bears a resemblance to a (renormalized) cosmological constant.

Note that the dominant term in $E_{Cas}(l)$ for $\mu l \ll 1$ coincides with the one corresponding to general Robin boundary conditions for the Sturm-Liouville problem subject to local boundary conditions (See \ref{local}). This can be explained by noticing that the most singular term of the Casimir energy for small $\mu l$ is due to the dependence of the eigenvalues $\omega_n$ on $n$ for $n\gg 1$, whose leading terms are independent of the $\beta$'s parameters (See Eq.~(\ref{21})). The third term in Eq.~(\ref{21}) do not depend neither on $\beta_1$ nor on $\beta_2$, showing that these parameters play no role in Eq.~(\ref{9}) up to this order in the asymptotic approximation to $\omega_n$.

Let us point out at this stage that the Casimir energy $E_{Cas}(l)$, which is, in principle, a measurable quantity, does not depend on the external mass scale introduced by the analytic regularization. Therefore, the \emph{constants} $\mathcal{E}_0$ and $\mathcal{E}_1$ \emph{run} with respect to the scale $\mu$. Indeed, taking into account that the last term in the right hand side of Eq.~(\ref{28}) does not depend on $\mu$ (See Eq.~(\ref{DELTA_E})),  
\begin{equation}\label{renorm-group}
    \mu \frac{\partial E_{Cas}}{\partial \mu}=0\quad \Rightarrow \quad
     \mu \frac{\partial \mathcal{E}_0}{\partial \mu}=\frac{\hbar \beta'_1}{2\pi \beta'_2} ,\quad
      \mu \frac{\partial \mathcal{E}_1}{\partial \mu}=-\frac{\hbar m^2}{4\pi},
\end{equation}
from which we get
\begin{equation}\label{running}
\begin{array}{c} \displaystyle
     \mathcal{E}_0(\mu)-\mathcal{E}_0(\mu_0)=\frac{\hbar \beta'_1}{2\pi \beta'_2}\log(\mu/\mu_0),
     \\ \\  \displaystyle
     \quad \mathcal{E}_1(\mu)-\mathcal{E}_1(\mu_0)=-\frac{\hbar m^2 }{4\pi}\log(\mu/\mu_0).
\end{array}
\end{equation}

\bigskip

In Appendix~(\ref{betap20}) we show that a similar iterative evaluation of the eigenvalues for the $\beta'_2=0,\ \beta'_1\neq 0$ case leads to the (unrenormalized) Casimir Energy
\begin{equation}\label{ECas-betap20}
    \begin{array}{c} \displaystyle
     E_Cas(l)=\frac{\hbar  \beta_2}{ \pi  \beta'_1
   \left(s+\frac{1}{2}\right)} + 
   \frac{\hbar  m^2 l}{2 \pi 
   \left(s+\frac{1}{2}\right)}
         \\ \\ \displaystyle
           - \frac{\pi  \hbar }{24 l} +
   \frac{\beta_2 \hbar  \log \left(\frac{\mu 
   l}{\pi }\right)}{2 \pi  \beta'_1} + O((\mu l)^0)
    \end{array}
\end{equation}
where again the (finite) most singular term for $\mu l\ll 1$ coincides with the singular behavior of the Casimir Energy for Dirichlet boundary conditions (See \ref{local}). In this case, the $\beta'_1$ term in the last line in Eq.~(\ref{9}) is dominant for large eigenvalues $\omega_n$, effectively imposing these boundary conditions. 

\section{Alternative evaluation of the meromorphic extension of $\zeta_A(s)$ for the massive case}

Note that $f(z)$ in Eq.~(\ref{14}) is an odd function which has no nonreal zeros, since $A$ is a positive self-adjoint operator. Moreover, its real zeroes $x_n>0,\, n\in \mathbb{N}$, correspond to the multiplicity one eigenvalues ${\omega_n}^2>m^2>0$ (The case $m=0$ will be considered later).

Then, employing the Cauchy's residue theorem, for real $s$ large enough and for an integration path encircling clockwise all the $x_n, n\in \mathbb{N}$,  we can rewrite Eq.~(\ref{20}) in the form of an integral on the complex plane,
\begin{equation}\label{30}
    \begin{array}{c}\displaystyle
      (\mu l)^{-2s} \zeta_A(s)=-\frac{1}{2\pi i}\oint_{-i\infty}^{i\infty} dz \left(z^2+M^2\right)^{-s} \frac{d}{dz} \log f(z)=
      \\ \\ \displaystyle
      =-\frac{M^{-2s}}{2\pi i}\oint_{-i\infty}^{i\infty} dz \left(z^2+1\right)^{-s} \frac{d}{dz} \log f(M z).
    \end{array}
\end{equation}
This can also be written as
\begin{equation}\label{30-1}
    \begin{array}{c}\displaystyle
      \left(\frac{m}{\mu}\right)^{2s}\zeta_A(s)=\frac{1}{\pi} \Im \left\{  e^{i\pi s}
      \int_1^\infty dy \left(y^2-1\right)^{-s} \frac{d}{dy} \log f(i M y)+
      \right.
      \\ \\ \displaystyle
      \left.
      +\lim_{\varepsilon \rightarrow 0^+}\int_\varepsilon^1 dy \left(1-y^2\right)^{-s}
      \frac{d}{dy} \log f(i M y) \right\}+
      \\ \\ \displaystyle
      -\frac{1}{2\pi i} \lim_{\varepsilon \rightarrow 0^+}\int_{-i }^{i } \varepsilon dz \left(\varepsilon^2 z^2+1\right)^{-s}
      \left[\frac{1}{\varepsilon z}+O(\varepsilon)\right],
    \end{array}
\end{equation}
where the last integral is evaluated on the half-circle $|z|=1, \Re z \geq 0$ and we can write
\begin{equation}\label{Logf}
    \log \left(2 i f(i M y)\right)=M y+\log (P(y))+\log \left(1-e^{-2 M y}\, \frac{ P(-y)}{P(y)}\right),
\end{equation}
with
\begin{equation}\label{Pdey}
    \begin{array}{c}
      P(y):=c-a M y-d M^2 y^2+b M^3 y^3=
      \\ \\ \displaystyle
      =l^2 \left\{ \left(\beta_1+ m^2 \beta_1'\right)-\left(\beta_2+m^2 \beta_2'\right) m y- {\beta_1'} m^2 y^2+ {\beta_2'} m^3 y^3\right\},
    \end{array}
\end{equation}
a real polynomial whose zeroes are independent of $l$.

The expression in the right hand side of Eq.~(\ref{Logf}) is useful to extract the asymptotic behavior of the integrand in the first line in Eq.~(\ref{30-1}). However, $P(y)$ could have zeros in the half-line $[1,\infty)$. In this case, the second and third terms in Eq.~(\ref{Logf})  present singularities that cancel each other. For simplicity, in the following we will assume that  $P(y)$  has no zeros in this half-line\footnote{Note that this do not significantly restrict the values of the parameters $\beta'_1$ and $\beta'_2$. See footnote \ref{restriction-zeros}}. If this were not the case, in the following analysis we should separate the integral in the first line in Eq.~(\ref{30-1}) into two parts, for $y\in [1,y_0]$ and $y\in [y_0,\infty)$, with $y_0$ a real number greater than the real zeros of this polynomial, and employ the expression in Eq.~(\ref{Logf}) only in the last region.

We remark that $P(y)$ is proportional to $l^2$ and the rational function
\begin{equation}\label{PsobreP}
   Q(y):= \frac{ P(-y)}{P(y)} \asymp \left\{
    \begin{array}{c} \displaystyle
      -1-\frac{2 {\beta_1'}}{  {\beta_2'} m y}+O\left({y^{-2}}\right), \quad {\rm if}\quad \beta_2' \neq 0,
      \\ \\ \displaystyle
       1-\frac{2 {\beta_2}}{{\beta_1'} m y} + O\left({y^{-2}}\right), \quad {\rm if}\quad \beta_2' = 0, \beta_1' \neq 0,
    \end{array}
    \right.
\end{equation}
is $l$-independent. In particular, notice that for $\beta'_2=0$ the condition $\rho= \beta_1' \beta_2 >0$ implies that $\beta_1'$ and $\beta_2$ have the same sign. Then, in this case $Q(y)$ is an increasing function for $y$ large enough.

So, we have
\bea\label{31}
      \frac{d}{dy} \log f(i M y)&=& M+ \frac{P'(y)}{P(y)}\nonumber\\ \\ \nonumber
      &&+
       \frac{P(y) P'(-y)+ \left[P'(y)+2 M P(y)\right]P(-y)}{P(y) \left[e^{2 M y} P(y)-P(-y)\right]}\in \mathbb{R},
\eea
where the last term is $O\left(e^{-2 M y}\right)$ and we have the asymptotic behaviors
\begin{equation}\label{32}
    \left\{
    \begin{array}{l} \displaystyle
    =M+\frac{3}{y}+\frac{\beta'_1}{ m \beta'_2 y^2}
    +O\left(y^{-3}\right)+
    O\left(e^{-2 M y}\right) , \,\, {\rm for} \quad y\gg 1, \beta'_2 \neq 0,
     \\ \\ \displaystyle
     =M+\frac{2}{y}-\frac{\beta_2}{ m \beta'_1 y^2}
   +O\left(y^{-3}\right)+
    O\left(e^{-2 M y}\right) , \,\, {\rm for} \quad y\gg 1, \beta'_2 = 0,
     \\ \\ \displaystyle
    = \frac{1}{y}+\frac{3a-6b-c+6d}{3 (a -c)}\, M^2 y +O\left(y^2\right), \quad {\rm for} \,\, y\ll 1.
    \end{array}
    \right.
\end{equation}

For any $\varepsilon>0$, the second integral on the right-hand side of Eq.~(\ref{30-1}) is convergent for $s<1$ and gives a real result. So, it can be discarded. The limit $\varepsilon\rightarrow 0^+$ of the last integral gives $\left(-\frac{1}{2}\right)$, and the contributions of $O\left(e^{-2My}\right)$ to the first integral also converge uniformly for $s<1$.

Therefore, from Eq.~(\ref{31}) and for $\frac{1}{2}<s<1$, we can write
\begin{equation}\label{zeta}
     \left(\frac{m}{\mu}\right)^{2s}\zeta_A(s)=I_1(s)+I_2(s)+F(s),
\end{equation}
where
\bea\label{33}
      &&I_1(s):=\frac{M}{\pi}\,\sin (\pi s) \int_1^\infty dy \left(y^2-1\right)^{-s}\nonumber\\
      &&=\frac{ M }{2 {\pi^{3/2} }}\, \sin (\pi s)  \Gamma (1-s) \Gamma \left(s-{1}/{2}\right),\nonumber\\
      &&I_2(s):= \frac{1}{\pi}\,\Im\left\{ e^{i \pi s} \int_1^\infty dy \left(y^2-1\right)^{-s}  \frac{d}{dy}\log P(y) \right\}\nonumber\\
      &&=\frac{1}{\pi}\,\Im\left\{ e^{i \pi s}  \int_1^\infty dy \left(y^2-1\right)^{-s}  \sum_{k=1}^3 \frac{1}{y-z_k} \right\},
\eea
with $z_k, k=1,2,3$ the zeroes of the cubic polynomial $P(y)$, and
\bea\label{33-1-F}
      F(s)&:=&-\frac{1}{2}+\frac{\sin(\pi s)}{\pi}\nonumber\\
      &&\times\int_1^\infty dy \left(y^2-1\right)^{-s} \frac{d}{dy} \log \left(1-e^{-2 M y}\, \frac{ P(-y)}{P(y)}\right) ,
\eea
which is analytic for $\Re s <1$ since the integrand is $O\left(e^{-2 M y}\right)$ for large $y$.

Finally, to determine the meromorphic extension of $\zeta_A(s)$ to $\Re s<1$ we must evaluate, for $\frac{1}{2}<s<1$, the integral $ I_2(s)$. Notice that the three zeros of $P(y)$ are all real or $z_1\in \mathbb{R}$ and $z_3={z_2}^*$, the complex conjugate of $z_2$, with $\Im z_2>0$. These poles in the integrand depend on $m$ and the parameters that determine the spectrum of $A$ in Eq.~(\ref{14}), but not on $l$. They satisfy\footnote{For the sake of clarity in the exposition, in deriving our results, we have simplified the analysis by assuming that $P(y)$ has no zeros on the half-line $[1,\infty)$. It is important to note that this assumption does not significantly restrict the possible values of the parameters $\beta'_1$ and $\beta'_2$. Specifically, if we assume, for example, that all three zeros are real and less than 1, then from Eq.~(\ref{zs}) it follows that
\[
{\beta_1'}<3 m \beta_2' ,\quad  {\beta_2}>-4 m^2 \beta_2', \quad  {\beta_1}> - m^3 \beta_2'- m^2 \beta_1'
\]
which is consistent with the parameter ranges specified in the first line of Eq.~(\ref{10}).\label{restriction-zeros}}
\begin{equation}\label{zs}
    \left\{
    \begin{array}{l}\displaystyle
      z_1+z_2+z_3=\frac{d}{b M}= \frac{\beta_1'}{m \beta_2' }
      \\ \\ \displaystyle
      z_1 z_2+z_1 z_3+z_2 z_3=\frac{-a}{b M^2}=-1-\frac{\beta_2}{m^2 \beta_2' }
      \\ \\ \displaystyle
      z_1 z_2 z_3= \frac{-c}{b M^3}=-\frac{\beta_1+m^2 \beta_1'}{ m^3 \beta_2'}
    \end{array}
    \right.
\end{equation}

\medskip

Now notice that, for $s<1$ and due to the factor $\Gamma \left(s-{1}/{2}\right)$ in the expression of $I_1(s)$, the analytic extension of its contribution to $\zeta_A(s)$ presents simple poles at $s=\frac{1}{2}-n$, for $n=0,1,2,\cdots$, with residue
\begin{equation}\label{33-1}
    {\rm Res}\left[ \left(\frac{m}{\mu }\right)^{-2 s} I_1(s)\right]_{s=\frac{1}{2}-n}=
    \frac{l \mu }{2 \pi ^{3/2} n!} \,\left(\frac{m}{\mu}\right)^{2 n} \Gamma \left(n+\frac{1}{2}\right).
\end{equation}

On the other hand, since we are assuming that the integrand in $I_2(s)$ is free of singularities for $y\in [1,\infty)$, it must be $\Re z_k <1$ or $\Im z_k \neq 0$. Taking into account that, for $u>0$, $-\pi<\alpha<\pi$ and $\frac{1}{2}<s<1$, we have \cite{Math}
\begin{equation}\label{35}
    \begin{array}{c}\displaystyle
      \int_1^\infty dy \, \frac{\left(y^2-1\right)^{-s}}{y+e^{i \alpha} u}=
    \frac{\pi }{\sin(2 \pi  s)}  \left[\left(e^{i \alpha } u\right)^2-1\right]^{-s}
    \\ \\ \displaystyle
    +\frac{1}{2 \sqrt{\pi }\, e^{i \alpha }u}\, \Gamma (1-s) \Gamma\left(s-\frac{1}{2}\right) \,
   _2F_1\left(\frac{1}{2},1;\frac{3}{2}-s;\frac{1}{\left(e^{ i \alpha}u\right)^2}\right),
    \end{array}
\end{equation}
where the hypergeometric function  (See {\bf 15.3.1} in \cite{AS}) 
\bea
 _2F_1\left(\frac{1}{2},1;\frac{3}{2}-s;\frac{1}{\left(e^{ i \alpha}u\right)^2}\right)\nonumber
  \\ =
 \left(\frac 12 -s \right)\int_0^1 (1-t)^{-1+\left(\frac 12 -s \right)}
 \left( 1-\frac{t}{\left(e^{ i \alpha}u\right)^2} \right) dt,
\eea
has a holomorphic extension for $\Re s<1$, 
we see that $I_2(s)$ can be written as
\begin{equation}\label{34}
    \begin{array}{c} \displaystyle
     I_2(s)= \frac{\sin (\pi s)}{\pi}  \sum_{k=1}^3\left\{
     \frac{\pi }{2\cos (\pi s)\sin( \pi  s)}  \left[{z_k}^2-1\right]^{-s}\right.
      \\ \\ \displaystyle
      \left.
      -\frac{1}{2 \sqrt{\pi }\, z_k}\, \Gamma (1-s) \Gamma\left(s-\frac{1}{2}\right) \,
   _2F_1\left(\frac{1}{2},1;\frac{3}{2}-s;\frac{1}{{z_k}^2}\right)
       \right\},
    \end{array}
\end{equation}
since the sum is manifestly real for real $s$. The meromorphic extension of the contribution of $I_2(s)$ to $\zeta_A(s)$ on the open half-plane $\Re s<1$ presents simple poles at $s=\frac{1}{2}-n$, with $n=0,1,2,\cdots$, with residues
\begin{equation}\label{36}
    \begin{array}{c}\displaystyle
      {\rm Res} \left[ \left(\frac{m}{\mu}\right)^{-2s} I_2(s)\right]_{s=\frac{1}{2}-n}
    = -\frac{1}{\pi }\left(\frac{m}{\mu }\right)^{2 n-1} \times
    \\ \\ \displaystyle
    \sum_{k=1}^3 \left\{ \frac{ (-1)^n}{2} \left({z_k}^2-1\right)^{n-\frac{1}{2}}+
    \frac{\Gamma \left(n+\frac{1}{2}\right) }{2 \sqrt{\pi}  n!\, z_k}\,_2F_1\left(\frac{1}{2},1;n+1;\frac{1}{{z_k}^2}\right)
    \right\}.
    \end{array}
\end{equation}
However, since the integrand in the definition of $I_2(s)$ in Eq.~(\ref{33}) behaves as
\bea\label{integ-asimpt}
          \left(y^2-1\right)^{-s}  \frac{d}{dy}\log P(y)&=&
      \left(y^2-1\right)^{-s}\nonumber\\
      &&\times  \left\{
    \begin{array}{c}  \displaystyle
     \frac{3}{y}+\frac{\beta_1'}{\beta_2'  m y^2}+O\left(y^{-3}\right) ,\quad \beta_2' \neq 0,
    \\ \\  \displaystyle
    \frac{2}{y}-\frac{\beta_2}{\beta_1' m y^2}+O\left(y^{-3}\right)  ,\quad \beta_2' = 0,
    \end{array}  \right.
\eea
the residue at $s=1/2$ vanishes. The first nonvanishing residue from $I_2(s)$ corresponds to $s=-1/2$:
\begin{equation}\label{res-12}
     {\rm Res} \left[ \left(\frac{m}{\mu}\right)^{-2s} I_2(s)\right]_{s=-\frac{1}{2}}
    =\left\{\begin{array}{c}\displaystyle
              -\frac{\beta_1'}{2 \pi   \mu  \beta_2'},\quad \beta_2' \neq 0,
              \\ \\  \displaystyle
              \frac{\beta_2}{2 \pi   \mu  \beta_1'} ,\quad \beta_2' = 0,
            \end{array}
    \right.
\end{equation}
independent of $l$ and $m$.

\medskip

Therefore, the analytic prolongation of $\zeta_A(s)$ is a meromorphic function in the half-plane $\Re s <1$ with isolated simple poles at the half-integers values of $s \leq 1/2$. Thus, $\zeta_A(s)$ shows a singularity structure which follows the general rule valid for the Sturm-Liouville operator subject to standard local  boundary conditions ({\it i.e.}\  dependent only on the values of the function and its {\it normal} derivative at the edge) \cite{Gilkey}, with isolated simple poles at $s=(d-n)/r \notin \mathbb{Z}_-$, where $d$ is the dimension of the manifold, $r$ is the order of the differential operator and $ n=0,1,2,\cdots$ 

The representation of $\zeta(s)$ as an integral over the complex plane in Eq.~(\ref{30}), fully equivalent to its series definition in Eq.~(\ref{20}), not only facilitates identifying its singular points but also provides expressions for its finite parts. As shown in Section~\ref{the_Casimir_energy}, which focuses on evaluating the Casimir energy of the model, these expressions are easier to analyze and evaluate as functions of $l$.

\section{The determinant of $A$}

According to the results in the previous Section, $\zeta_A(s)$ is analytic in a neighborhood of the origin. Indeed, from Eq.~(\ref{zeta}), the first line in Eq.~(\ref{33}), and Eqs.~(\ref{33-1}) and (\ref{34}), we get for $s\approx 0$
\begin{equation}\label{39}
    \begin{array}{c}\displaystyle
                \zeta_A(s)= -M s + \left[ \frac{3}{2} - \left(3 \log \left(\frac{m}{\mu }\right)+ \sum_{k=1}^3 \log(z_k-1)\right) s \right]
     \\ \\ \displaystyle
     +\left[1-2s \log \left(\frac{m}{\mu }\right)  \right]
     \left\{  -\frac{1}{2}+s F'(0) \right\} + O\left(s^2\right)
    \end{array}
\end{equation}
and, from the usual definition of the functional determinant \cite{Ray-Singer,Dowker,Hawking},
\begin{equation}\label{39-Det}
    \begin{array}{c}\displaystyle
      \log {\rm Det}\left( A/\mu^2\right):=-{\zeta_A}'(0)
      \\ \\ \displaystyle
      =m l +2 \log \left(\frac{m}{\mu }\right)+ \sum_{k=1}^3 \log(z_k-1) - F'(0),
    \end{array}
\end{equation}
where
\begin{equation}\label{logs}
    \sum_{k=1}^3 \log(z_k-1)=\log\left\{ \frac{-P(1)}{\beta_2' l^2 m^3} \right\}
\end{equation}
and
\begin{equation}\label{Fprima}
    F'(0)=-\log \left[1-e^{-2 m l}\frac{P(-1)}{P(1)}\right].
\end{equation}
Notice that ${\zeta_A}(0)=1$ and ${\rm Det} \left( A/\mu^2\right)$ do depend on the external scale $\mu$.

Similar considerations apply for the $\beta'_2=0$ case.

\section{The Casimir energy}\label{the_Casimir_energy}

As previously remarked, $\zeta_A(s)$ has a simple pole at $s=-1/2$. From Eq.~(\ref{33-1}) with $n=1$ and Eq.~(\ref{res-12}), we get
\begin{equation}\label{40}
              \zeta_A(s) = \left\{\frac{l  m^2 }{4 \pi \mu} -\frac{\beta_1'}{2 \pi   \mu \beta_2'} \right\} \frac{1}{\left(s+\frac{1}{2}\right)} + O\left(\left(s+{1}/{2}\right)^0\right).
\end{equation}
Note that the singular term coincides with that of Eq.~(\ref{27}).

Around $s=-1/2$ we have
\bea\label{desarrollo1}
    \left(\frac{m}{\mu}\right)^{-2s} I_1(s)&=&\frac{l m^2}{4 \pi  \mu  } \Bigg\{\frac{1}{\left(s+\frac{1}{2}\right)} -
    \left[2 \log \left(\frac{m}{2\mu}\right)+1\right]\nonumber\\
    &&+O\left(s+\frac{1}{2}\right)\Bigg\}
\eea
and
\begin{equation}\label{desarrollo2}
    \begin{array}{c} \displaystyle
      \left(\frac{m}{\mu}\right)^{-2s} I_2(s)= \frac{3m}{2\mu} -\frac{\beta'_1}{2 \pi \mu  \beta'_2  } \left\{
      \frac{1}{s+\frac{1}{2}} -2 \left[\log \left(\frac{m}{2 \mu }\right)+1\right]\right\}+
       \\ \\  \displaystyle
       -\frac{m}{\pi \mu} \int_1^\infty dy \sqrt{y^2-1} \left\{ \frac{P'(y)}{P(y)}  -\frac{3}{y}-\frac{ \beta_1' }{\beta_2' m  y^2} \right\}
       +O\left(s+\frac{1}{2}\right),
    \end{array}
\end{equation}
where the integral in the last line can also be written interms of the zeroes of $P(y)$ as
\begin{equation}\label{desarrollo3}
    -\frac{m}{\pi \mu} \lim_{\Lambda \rightarrow \infty}\int_1^\Lambda dy \sqrt{y^2-1} \left\{ \sum_{k=1}^3\frac{1}{y-z_k}  -\frac{3}{y}-\frac{\beta_1'}{\beta_2' m  y^2} \right\}
\end{equation}

Taking into account that, for $z\in \mathbb{C}$, the primitive
\bea\label{primitiva}
      \int \frac{\sqrt{y^2-1}}{y-z}\, dy &=&
      \sqrt{y^2-1}+ z \log \left(y+\sqrt{y^2-1}\right)\nonumber\\
      &&-\sqrt{1-z^2}\,  {\rm arctan}\left(\frac{\sqrt{y^2-1} \sqrt{1- z^2}}{1- y z}\right)
\eea
vanishes for $y=1$ and, for $y=\Lambda \gg 1$, behaves as
\bea\label{largeLambda}
    \int \frac{\sqrt{y^2-1}}{y-z }\, dy &=&
    \Lambda +z \left(\log \Lambda +\log 2 \right)\nonumber\\
    &&+\sqrt{1-z^2}\,  {\rm arctan}\left(\frac{\sqrt{1-z^2}}{z}\right)+ O\left(\Lambda^{-1}\right),
\eea
we can write
\bea\label{I2demenos1medio}
      &&\left(\frac{m}{\mu}\right)^{-2s} I_2(s)= \frac{3m}{2\mu} -\frac{\beta'_1}{2 \pi \mu  \beta'_2  }   \left\{
      \frac{1}{s+\frac{1}{2}} -2 \left[\log \left(\frac{m}{2 \mu }\right)+1\right]\right\}\nonumber\\
       &&-\frac{m}{\pi \mu} \left\{\frac{3 \pi}{2}+\frac{ \beta_1' }{\beta_2' m} + \sum_{k=1}^3 \sqrt{{1-{z_k}^2}}\, {\rm arctan}\left(\frac{\sqrt{{1-{z_k}^2}}}{z_k}\right) \right\}\nonumber\\
       &&+O\left(s+\frac{1}{2}\right)\nonumber\\
       &&= -\frac{\beta_1'}{2 \pi  \mu \beta_2'} \left\{
      \frac{1}{s+\frac{1}{2}} -2 \log \left(\frac{m}{2 \mu }\right)\right\}\nonumber\\
      &&-\frac{m}{\pi \mu}  \sum_{k=1}^3 \sqrt{{1-{z_k}^2}}\, {\rm arctan}\left(\frac{\sqrt{{1-{z_k}^2}}}{z_k}\right)+O\left(s+\frac{1}{2}\right)
\eea
where use has been made of Eq.~(\ref{zs}).

\medskip

Taking into account that the zeroes of $P(y)$ are $l-$independent, after appropriate subtractions of the divergent linear in $l$ and constant terms, thus renormalizing the linear energy density and the zero energy level, (and assuming that $P(y)$ has no zeroes in $[1,\infty)$) we can define de Casimir energy as
\begin{equation}\label{41}
    \begin{array}{c} \displaystyle
      E_{Cas}(l):=\left[ \mathcal{E}_0 -\frac{\hbar \beta'_1}{2\pi \beta'_2}\log\left(\frac \mu m\right)\right]+
      \\ \\ \displaystyle
      + l\left[\mathcal{E}_1 +\frac{\hbar m^2}{4\pi} \log\left(\frac \mu m\right)\right]  + \Delta E(l),
      \\ \\ \displaystyle
      \Delta E(l)= -\frac{\hbar m}{2\pi}
      \int_1^\infty dy \sqrt{y^2-1}\, \frac{d}{dy} \log \left(1-e^{-2 M y}\, \frac{ P(-y)}{P(y)}\right),
    \end{array}
\end{equation}
where we have explicitly retained $\mu$-dependent constant and linear terms and use has been made of the definition of $F(s)$ given in Eq.~(\ref{33-1-F}). After an integration by parts, $\Delta E(l)$ reads as
\be \label{DeltaECas}
\Delta E(l):= \frac{\hbar m}{2\pi}
      \int_1^\infty dy \, \frac{y}{\sqrt{y^2-1}}\, \log \left(1-e^{-2 M y} Q(y)\right) ,
\ee
where we have defined the rational function $Q(y):= \frac{ P(-y)}{P(y)}$. As discussed in  Section~(\ref{spectrum}), $\mathcal{E}_0$ and $\mathcal{E}_1$ are phenomenological parameters: $\mathcal{E}_0$ redefines the energy reference level, while $\mathcal{E}_1$ represents a renormalized energy density governing the system's behavior at large $l$. It follows from Eq.~(\ref{41}) that both parameters run with $\mu$ to ensure the independence of $E_{Cas}(l)$ from the external mass scale, as discussed at the end of Section \ref{spectrum}. Eq.~(\ref{41}) should be compared with Eq.~(\ref{28}).

\subsection{The Casimir energy as a function of $l$} \label{ECasimir}

Eq.~(\ref{DeltaECas}) offers an adequate expression for the numerical evaluation of $\Delta E(l)$, except possibly for $M=m l \ll 1$, region where we give some estimates.  

The large-$l$ asymptotic behavior of $E_{Cas}(l)$  is dominated by the (undetermined) linear term, since $ \Delta E(l)$ in Eq.~(\ref{41}) is $O\left(e^{-2m l}\right)$,
\begin{equation}\label{Cas-1}
    \begin{array}{c} \displaystyle
     \Delta E(l)= E_{Cas}(l)-\left\{\mathcal{E}_0+ \mathcal{E}_1 l\right\}  
      \\ \\ \displaystyle
       \approx -\frac{\hbar m}{2\pi} \, e^{-2 M}
      \int_1^\infty dy \, \frac{y}{\sqrt{y^2-1}}\, e^{-2 M (y-1)} Q(y)
      + O\left(e^{-4 M}\right).
    \end{array}
\end{equation}

\medskip

On the other hand, for small $l$ (small $M=m l$), two cases must be considered, $\beta'_2 \neq 0$ and $\beta'_2 = 0$ according to the asymptotic behavior of $Q(y)$ shown in Eq.~(\ref{PsobreP}), which will be shown to present completely different behaviors.

\smallskip

Let us first consider the case $\beta'_2 \neq 0$, for which $\lim_{y \rightarrow \infty} Q(y)= -1$.  For a given $\varepsilon>0$, let $y_0$ be large enough\footnote{$y_0> \frac{1}{\varepsilon}\left|\frac{2\beta'_1}{m \beta_2}\right|$ (See Eq.~(\ref{PsobreP})).} to have $-1-\varepsilon < Q(y) < -1+\varepsilon$, $\forall y>y_0$. Therefore,
\bea\label{cotas1}
      \log\left\{1+(1-\varepsilon)e^{-2M y} \right\} &<& \log\left\{ 1-e^{-2M y} Q(y)\right\}\nonumber\\ 
      &<& \log\left\{1+(1+\varepsilon)e^{-2M y} \right\}
\eea
since the logarithm is a monotonously increasing function of its argument.

Taking into account that  (See Eq. 4.1.33 in \cite{AS})
\begin{equation}\label{cotas2}
    \frac{x}{1+x} < \log(1+x)<x, \quad {\rm for}\ x>-1, \ x\neq 0,
\end{equation}
we can write from Eq.~(\ref{cotas1})
\begin{equation}\label{cotas3}
    \frac{(1-\varepsilon) e^{-2M y}}{1+(1-\varepsilon) e^{-2M y}} < \log\left\{ 1-e^{-2M y} Q(y)\right\} <(1+\varepsilon)  e^{-2M y},
\end{equation}
and then,
\bea\label{cotas4}
       \int_{y_0}^\infty  \frac{dy \, y}{\sqrt{y^2-1}}\, \frac{(1-\varepsilon) e^{-2M y}}{1+(1-\varepsilon) e^{-2M y}} 
     &<&\int_{y_0}^\infty \frac{dy \, y}{\sqrt{y^2-1}}\, \log\left\{ 1-e^{-2M y} Q(y)\right\}\nonumber\\  
      &<& (1+\varepsilon)        \int_{y_0}^\infty dy \, \frac{ y}{\sqrt{y^2-1}}\,e^{-2M y} ,
\eea
since the integrands take positive values.

Now, for the integral in the right hand side of Eq.~(\ref{cotas4}), by integrating by parts we get
\bea\label{cotas5}
     &&-{\sqrt{y_0^2-1}}\, e^{-2M y_0}+2M \int_{y_0}^\infty  dy {\sqrt{y^2-1}}\, e^{-2M y}<
   2M \int_{y_0}^\infty  dy \,  y \, e^{-2M y}\nonumber\\
   &&=  -M \frac{\partial}{\partial M} \int_{y_0}^\infty  dy \, e^{-2M y}\nonumber\\
   &&=  -M \frac{\partial}{\partial M}  \left(\frac{e^{-2M y_0}}{2M} \right)=\frac{1}{2M} + O\left(M^0\right).
\eea

For the integral in the left hand side of Eq.~(\ref{cotas4}), taking into account that $y>{\sqrt{y^2-1}}$, we can write
\bea\label{cotas6}
    &&\int_{y_0}^\infty dy \, \frac{ y}{\sqrt{y^2-1}}\, \frac{(1-\varepsilon) e^{-2M y}}{1+(1-\varepsilon) e^{-2M y}}  >
    \int_{y_0}^\infty dy\, \frac{(1-\varepsilon) e^{-2M y}}{1+(1-\varepsilon) e^{-2M y}}\nonumber\\  &&-\frac{1}{2M} \int_{y_0}^\infty dy\,\frac{d}{dy} \, \log\left( 1+(1-\varepsilon)e^{-2M y} \right)\nonumber\\
     &&= \frac{1}{2M} \, \log\left( 1+(1-\varepsilon)e^{-2M y_0} \right)
    =\frac{\log(2-\varepsilon)}{2M}+ O\left(M^0\right).
\eea

Moreover,
\begin{equation}\label{cotas7}
    \begin{array}{c} \displaystyle
    \int_1^{y_0} \frac{dy \, y}{\sqrt{y^2-1}}\, \log \left(1-e^{-2 M y} Q(y)\right)=
     \\ \\ \displaystyle
    =\int_1^{y_0} \frac{dy \, y}{\sqrt{y^2-1}}\, \log \left(1- Q(y)\right)+ O\left(M\right).
    \end{array}
\end{equation}

\begin{figure}
    \begin{tikzpicture}
    \node(a){\includegraphics[width=0.8\textwidth]{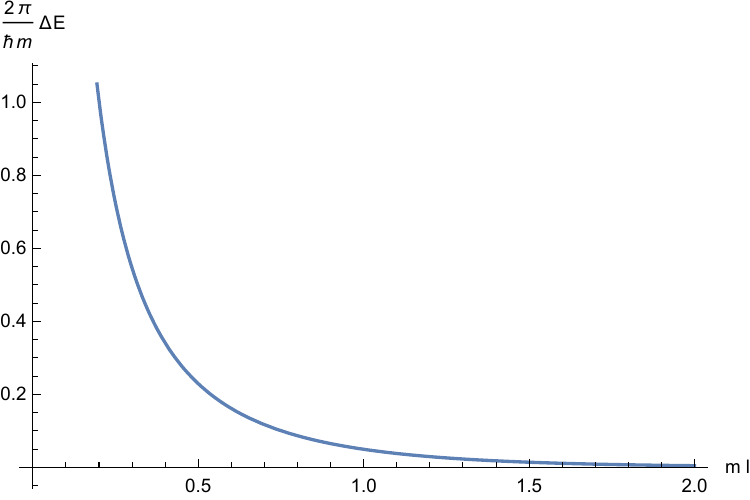}};
    \node at (a.north east)
    [
    anchor=center,
    xshift=-35mm,
    yshift=-35mm
    ]
    {
        \includegraphics[width=0.45\textwidth]{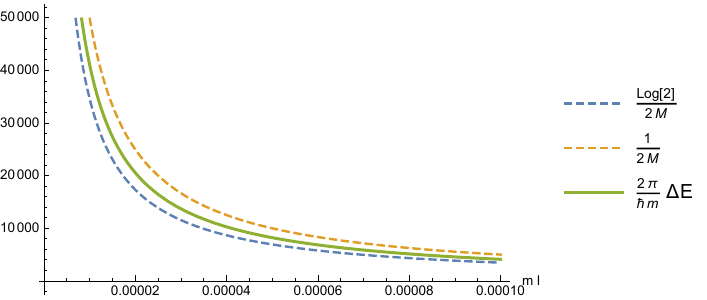}
    };
    \end{tikzpicture}
     \caption{$\Delta E$ as a function of $M=m l$, with $ b_1=1/2, b_2=-1, b'_1=-1/2, b'_2=1/2$.}
     \label{Ecas1}
\end{figure}

\begin{figure}
    \begin{tikzpicture}
    \node(a){\includegraphics[width=0.8\textwidth]{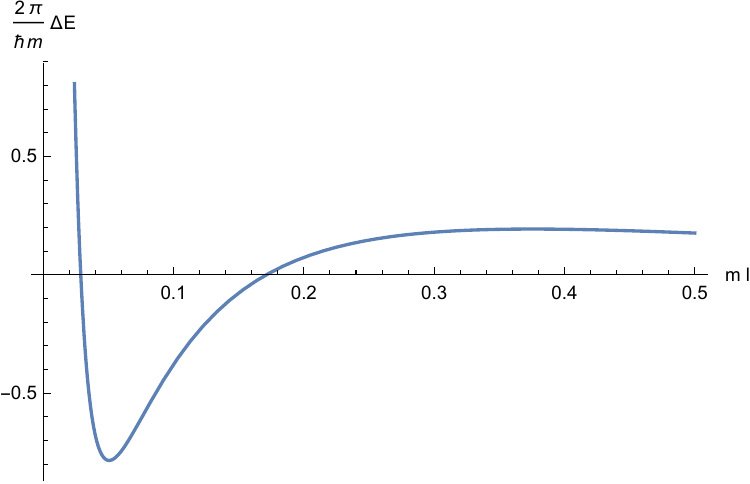}};
    \node at (a.north east)
    [
    anchor=center,
    xshift=-35mm,
    yshift=-15mm
    ]
    {
        \includegraphics[width=0.45\textwidth]{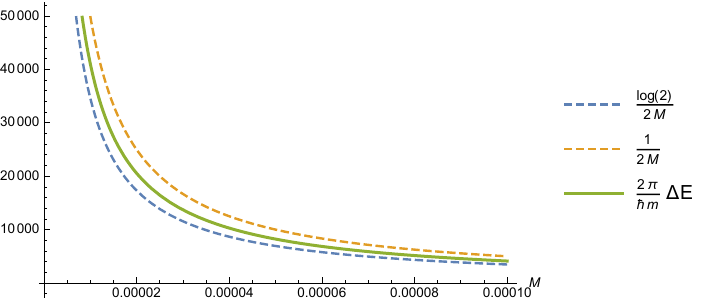}
    };
    \end{tikzpicture}
    \caption{$\Delta E$ as a function of $M=m l$ for $ b_1=1, b_2=-150, b'_1=-130, b'_2=7$.}
     \label{Ecas2}
\end{figure}

Therefore, for $\beta'_2\neq 0$ and $M \ll 1$, the asymptotic contribution from $\Delta E(l)$ defined in Eq.~(\ref{DeltaECas}) (up to the factor $ \frac{\hbar m}{2\pi}$) presents a singular behavior bounded as\footnote{Notice that these bounds are consistent with Eq.~(\ref{28}) since $\frac{\log 2}{4\pi }<\frac{\pi }{48}<\frac{1}{4\pi }$.}
\begin{equation}\label{cotas8}
    \frac{\log(2-\varepsilon)}{2M}+ O\left(M^0\right) <
     \frac{2\pi}{\hbar m}\, \Delta E(l)
    < \frac{1+\varepsilon}{2M}+ O\left(M^0\right)
\end{equation}
for any $\varepsilon>0$ and independently of the values of the parameters defining the spectral dependent boundary condition in Eq.~(\ref{9}).  Figures~(\ref{Ecas1}) and (\ref{Ecas2}) show two possible behaviors of the Casimir energy in this case: The first corresponds to repulsion at all distances, while the second exhibits repulsive behavior at large distances, followed by an attractive effective potential with a local minimum in an intermediate region, before becoming repulsive again at short distances as described in Eq.~(\ref{cotas8}). Note that the existence of a root of $P(-y)$ in the half-line $(1,\infty)$ allows for these local extrema of the Casimir energy. In these figures, $E_{Cas}$ has been plotted as a function of $M=m l$, and the parameters $\beta$ have been scaled with the power of $m$ corresponding to their dimension: $\beta_1=b_1 m^2,  \beta_2=b_2 m,  \beta'_1=b'_1 m^0, \beta'_2=b'_2/m$, with $b_i$ pure numbers.

\smallskip

Let us now consider the case $\beta'_2=0$, for which $Q(y)$ is an increasing function for $y\gg 1$ with $\lim_{y \rightarrow \infty} Q(y)= 1$. So, for a given $\varepsilon>0$, let $y_0$ be large enough\footnote{$y_0>\frac{1}{\varepsilon}\left|\frac{2\beta_2}{m\beta'_1}\right|$ (See 
Eq.~(\ref{PsobreP})).} to have $1-\varepsilon < Q(y)<1$  if $y>y_0$.  Then,
\begin{equation}\label{cotas9}
          \log\left(1- e^{-2M y}\right) < \log\left(1- e^{-2M y} Q(y)\right) < \log\left(1-(1-\varepsilon) e^{-2M y}\right).
\end{equation}

Taking into account that (See Eq. 4.1.34 in \cite{AS})
\begin{equation}\label{cotas10}
    \frac{-x}{1-x} < \log(1-x) < -x, \quad {\rm for} \ x<1, x\neq 0,
\end{equation}
we can write
\bea\label{cotas11}
      &&\int_{y_0}^\infty \frac{dy \, y}{\sqrt{y^2 -1}} \left(\frac{- e^{-2M y}}{1- e^{-2M y}}\right)<
    \int_{y_0}^\infty \frac{dy \, y}{\sqrt{y^2 -1}}\, \log\left(1- e^{-2M y} Q(y)\right)\nonumber\\
      &&< - (1-\varepsilon) \int_{y_0}^\infty \frac{dy \, y}{\sqrt{y^2 -1}}\,  e^{-2M y} <  -  (1-\varepsilon) \int_{y_0}^\infty dy\,  e^{-2M y}\nonumber\\
      &&<- \frac{(1-\varepsilon)}{2M}+ O\left(M^0\right).
\eea

For the integral in the left hand side of \ref{cotas11}, we have
\begin{equation}\label{cotas12}
    \begin{array}{c} \displaystyle
       \int_{y_0}^\infty \frac{dy \, y}{\sqrt{y^2 -1}} \left(\frac{ e^{-2M y}}{1- e^{-2M y}}\right)<
        \int_{1}^\infty d {\sqrt{y^2 -1}} \frac{1 }{\left(e^{2M y}-1\right) }
       \\ \\ \displaystyle
      =-  \int_{1}^\infty dy\,  {\sqrt{y^2 -1}} \left(\frac{-2M e^{2M y}}{\left(e^{2M y}-1\right)^2 }\right)
      \\ \\ \displaystyle
      <  M \int_{1}^\infty dy \left(\frac{2y e^{2M y}}{\left(e^{2M y}-1\right)^2 }\right)=
      -  M \frac{\partial}{\partial M} \int_{1}^\infty d y \frac{1 }{\left(e^{2M y}-1\right) }=
      \\ \\ \displaystyle
      -  M \frac{\partial}{\partial M} \int_{1}^\infty  \frac{dy}{2M} \frac{d}{dy} \log\left({1-e^{-2M y}}\right)
        M \frac{\partial}{\partial M} \left\{\frac{\log\left({1-e^{-2M}}\right)}{2M}  \right\}
        \\ \\ \displaystyle
        =\frac{1}{e^{2M}-1}-\frac{\log\left(1-e^{-2M}\right)}{2M}= \frac{1-\log(2M)}{2M}+O(M).
    \end{array}
\end{equation}

\begin{figure}
    \begin{tikzpicture}
    \node(a){\includegraphics[width=0.8\textwidth]{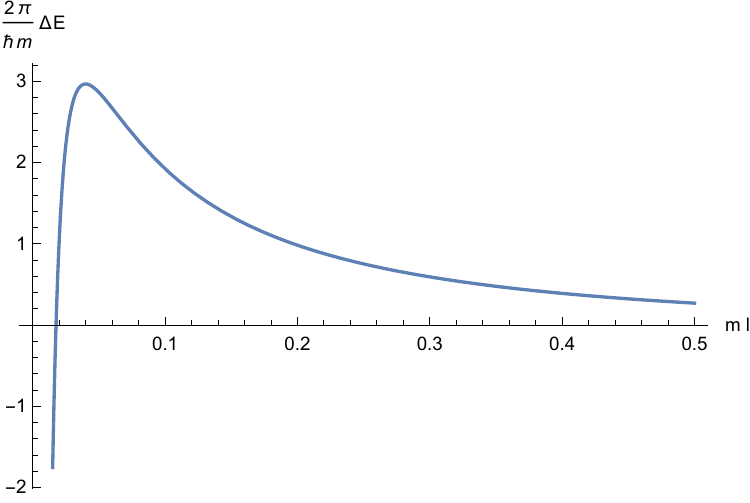}};
    \node at (a.north east)
    [
    anchor=center,
    xshift=-35mm,
    yshift=-25mm
    ]
    {
        \includegraphics[width=0.45\textwidth]{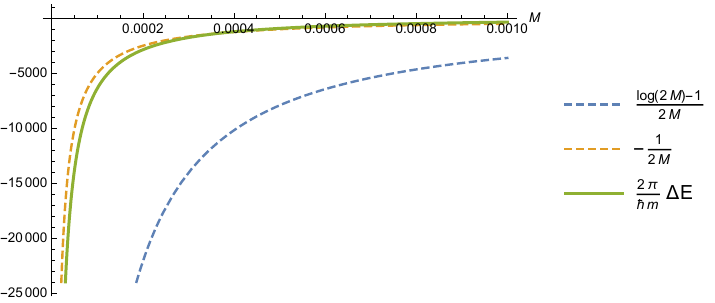}
    };
    \end{tikzpicture}
     \caption{$\Delta E$ as a function of $M= m l$ for $ b_1=10, b_2=-22, b'_1=-1, b'_2=0$.}
     \label{Ecas3}
\end{figure}

\begin{figure}
    \begin{tikzpicture}
    \node(a){\includegraphics[width=0.8\textwidth]{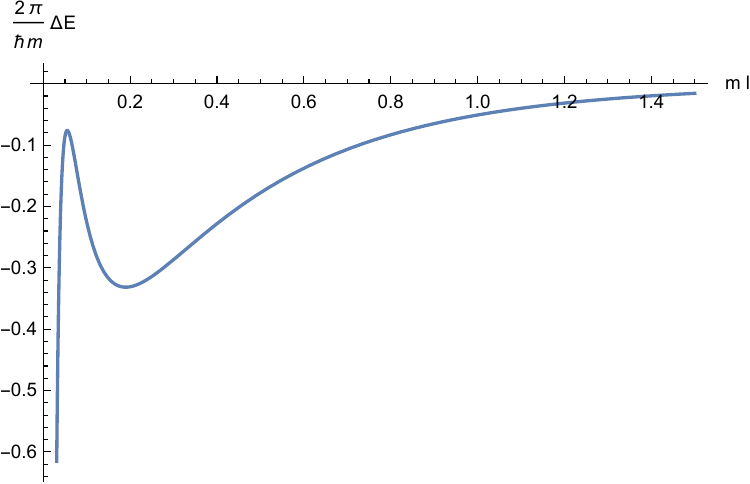}};
    \node at (a.north east)
    [
    anchor=center,
    xshift=-35mm,
    yshift=-55mm
    ]
    {
        \includegraphics[width=0.45\textwidth]{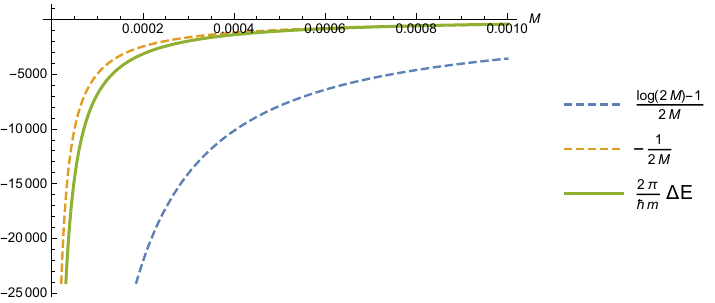}
    };
    \end{tikzpicture}
    \caption{$\Delta E$ as a function of $M= m l$ for $ b_1=61, b_2=-23, b'_1=-1, b'_2=0$.}
     \label{Ecas4}
\end{figure}

Therefore, taking into account Eq.~(\ref{cotas7}), for $\beta'_2 = 0$ and $M \ll 1$, the asymptotic contribution from $\Delta E(l)$ to the Casimir energy presents also a singular behavior, but this time bounded as\footnote{These bounds are also consistent with Eq.~(\ref{ECas-betap20}) since $\frac{-1+\log[M]}{2\pi }<\frac{-\pi }{24}<\frac{-1}{4\pi }$ is satisfied for $M<1$, even though the lower bound found deviates significantly.}
\begin{equation}\label{cotas13}
    \frac{-1+\log(2M)}{2M}+O(M^0)< \frac{2\pi}{\hbar m}\, \Delta E(l) \leq  \frac{-1}{2M}+ O\left(M^0\right)
\end{equation}
independently of the values of the other parameters. Figures (\ref{Ecas3}) and (\ref{Ecas4}) show two possible behaviors for this case: The first corresponds to a repulsive potential for large distances, with a local maximum and a strong singular attraction at shorter distances bounded as in Eq.~(\ref{cotas13}). The second one corresponds to an attractive potential for large distances with a local minimum, and a strong attraction for distances below a local maximum. Again, this is a consequence of the existence of zeros of $P(-y)$ that lie in the half-line $(1,\infty)$.

\section{The massless case}

With $m=0$, the solution of Eq.~(\ref{11}) is given by
\begin{equation}\label{m-1}
   \varphi_1(z) \sim \sin \left( z \omega\right),
\end{equation}
where the eigenvalues $\omega$ satisfy
\begin{equation}\label{m-2}
    \left( \beta_2+\omega^2\beta_2'\right)  \omega \cos \left( l \omega \right)=
      \left( \beta_1 +\omega^2 \beta_1' \right)  \sin \left( l \omega\right).
\end{equation}
Notice that $\omega=0$ is not an eigenvalue. So, the first one is positive.

By defining  $x:=\omega/\mu$, where $\mu$ is the external mass scale, and calling
\begin{equation}\label{m-3}
    g(x):=  \left( \beta_2+\beta_2'\mu^2 x^2\right)  \mu x \cos \left(  \mu l x \right) -
      \left( \beta_1 +\beta_1'\mu^2  x^2 \right)  \sin \left( \mu l x \right),
\end{equation}
the $\zeta$-function of this system can be written as
\begin{equation}\label{m-4}
   \zeta_A(s)= \sum_{n=1}^\infty x_n^{-2s} =
    \frac{1}{2\pi i } \oint_{\mathcal{C}} dz\, z^{-2s} \frac{d}{dz}\log g(z),
\end{equation}
where $\mathcal{C}$ encircles counterclockwise the spectrum of $A$, which is contained in the half-line $z>0$. For $\Re s>1/2$, Cauchy's theorem allows to write
\begin{equation}\label{m-5}
     \zeta_A(s)=-  \frac{1}{2\pi i }  \left\{ \int_{-i \infty}^{-i}+  \int_{\mathcal{C}_0}+  \int_{i}^{i \infty}+ \right\}dz\, z^{-2s} \frac{d}{dz}\log g(z),
\end{equation}
where $\mathcal{C}_0$ lies on the open right half-plane, going from $-i$ to $i$ and crossing the real axis at a positive real less than the first zero of $g(x)$.

Taking into account that $g( e^{-i \pi} z)=e^{-i \pi}   g(z)$ and $g(z)= \mu(\beta_2-l \beta_1) z + O \left(z^3\right)$, we have
\bea\label{m-6}
      &&\zeta_A(s)= \frac{\sin (\pi s)}{\pi} \int_1^\infty dy\, y^{-2s} \frac{d}{dy}\log g(i y)\\
       &&-\frac{1}{2\pi i }   \int_{-i}^{i}dz\, z^{-2s}  \frac{d}{dz}\log \left(\frac{g(z)}{z}\right)
        -\frac{1}{2\pi i }   \int_{-i}^{i}dz\, z^{-2s-1}\nonumber\\  
        &&= \frac{\sin (\pi s)}{\pi}\Bigg\{  \int_1^\infty dy\, y^{-2s} \frac{d}{dy}\log g(i y)\nonumber\\
        &&+
        \int_0^1 dy\, y^{-2s} \frac{d}{dy}\log\left(\frac{g(i y)}{y}\right) - \frac{1}{2s} \Bigg\},\nonumber
\eea
for $\frac{1}{2}<s<1$, region from which $ \zeta_A(s)$ can be analytically continued. Notice that the second integral into the brackets is analytic for $\Re s<1$, while the last term presents a simple pole at $s=0$ which cancels out against the zero of $\sin \pi s$. All the singular points of $ \zeta_A(s)$ for $\Re s<1$ are due to the first integral into these brackets.

Indeed,  we have
\begin{equation}\label{m-8}
    \begin{array}{c} \displaystyle
   2 i\,  g(i y)= e^{ \mu l y } p(y)- e^{-\mu l y } p(-y) \quad \Rightarrow
      \\ \\ \displaystyle
         \frac{d}{dy} \log \left[2 i  g(i y)\right]= \mu l + \frac{p'(y)}{p(y)} + \frac{d}{dy}\log\left[1- e^{-2\mu l y } q(y)\right],
    \end{array}
\end{equation}
where
\begin{equation}\label{m-9}
    p(y):= \beta_2' \mu^3 y^3 - \beta_1' \mu^2 y^2 -\beta_2 \mu y+ \beta_1,
    \quad
    q(y):=  \frac{p(-y)}{p(y)},
\end{equation}
with $p(y)$ independent of $l$. Then, the contribution of the first term in the second line of Eq.~(\ref{m-8}) to $\zeta_A(s)$ is given by
\begin{equation}\label{m-10}
    \frac{\mu l \sin (\pi s)}{\pi}  \int_1^\infty dy\, y^{-2s}  = \frac{\mu l}{2\pi}\, \frac{\sin \pi s}{s-1/2},
\end{equation}
which introduces a simple pole at $s=1/2$.

For the contribution of the second term, notice that
\bea\label{m-11}
    \frac{d}{dy}\log p(y) =   \left\{
    \begin{array}{l}\displaystyle
    \frac{3}{y}+\frac{{\beta_1'}}{{\beta_2'} \mu  y^2}+\frac{2 {\beta_2}
   {\beta_2'}+{\beta_1'}^2}{{\beta_2'}^2 \mu ^2  y^3}+O\left(y^{-4}\right),\,\, {\rm if} \ \beta_2' \neq 0,\\ \\
      \frac{2}{y}-\frac{{\beta_2}}{{\beta_1'} \mu  y^2}+\frac{2 {\beta_1}
   {\beta_1'}+{\beta_2}^2}{{\beta_1'}^2 \mu ^2  y^3}+O\left(y^{-4}\right) , \,\, {\rm if}\ \beta_2'=0, \beta_1'\neq 0.
    \end{array}
    \right.
\eea
In both cases, the successive contributions to  $\zeta_A(s)$ are proportional to integrals of the form
\begin{equation}\label{m-12}
      \frac{\sin (\pi s)}{\pi} \int_1^\infty dy\, y^{-2s-n} =   \frac{\sin (\pi  s)}{2\pi  \left(s-\frac{1-n}{2}\right)},
\end{equation}
introducing simple poles at even values of $n$, $n= 2,4,6,\cdots$

This can also be seen by expressing the left hand side of Eq.~(\ref{m-11}) in terms of the zeros of $p(y)$ (which are independent of $l$),
\begin{equation}\label{m-13}
    p(y)= \left\{
    \begin{array}{c} \displaystyle
    \beta_2' \mu^3 (y-u_1)(y-u_2)(y-u_3),\quad {\rm if} \ \beta_2'\neq 0,
      \\ \\ \displaystyle
      -\beta_1' \mu^2 (y-v_1)(y-v_2), \quad {\rm if}\  \beta_2'=0, \beta_1'\neq 0,
    \end{array}
    \right.
\end{equation}
where
\begin{equation}\label{m-14}
    u_1+u_2+u_3=\frac{\beta_1'}{\beta_2' \mu}.\quad u_1 u_2+u_1 u_3+u_2 u_3= -\frac{\beta_2}{\beta_2' \mu^2},
    \quad u_1 u_2 u_3= -\frac{\beta_1}{\beta_2' \mu^3},
\end{equation}
or
\begin{equation}\label{m-15}
    v_1+v_2=\frac{\beta_2}{\beta_1' \mu},\quad v_1 v_2=\frac{\beta_1}{\beta_1' \mu^2}
\end{equation}
respectively. Then, for $\Re s>0$,  this contribution has the form
\begin{equation}\label{m-16}
    \begin{array}{c} \displaystyle
    \frac{\sin (\pi s)}{\pi} \int_1^\infty dy\, y^{-2s} \frac{d}{dy}\log p(y)=
     \frac{\sin (\pi s)}{\pi}  \sum_{k}  \int_1^\infty dy\, \frac{y^{-2s} }{y-\mathfrak{z}_k}
      \\ \\ \displaystyle
      = \frac{\sin (\pi s)}{\pi}  \sum_{k}
      \left\{ \frac{1}{2s}+\frac{\mathfrak{z}_k}{2s+1}
        +\frac{\mathfrak{z}_k^2}{2s+2} \, _2F_1(1,2 s+2;2 s+3;\mathfrak{z}_k)   \right\} 
    \end{array}
\end{equation}
where $\mathfrak{z}_k$ are the zeroes of $p(y)$ in each case and the nearest poles of the meromorphic extension are explicitly shown in the last line.

Finally, the last term in the second line of Eq.~(\ref{m-8}) is exponentially vanishing for $y \rightarrow \infty$, and its contribution to $\zeta_A(s)$ is analytic for $\Re s<1$ (as long as $p(y)$ has no zeroes for $y\in [1,\infty)$, as we will assume for definiteness\footnote{This does not significantly restrict the possible values of the parameters in the boundary condition. Specifically, from the first line of Eq.~(\ref{m-14}) and in the case of three real negative zeros we get $\beta_1>0$, $\beta'_1,\beta_2<0$ for $\beta'_2>0$, in agreement with Eq.~(\ref{10}). On the other hand, for $\beta'_2=0$ and $\beta'_1<0$, the discriminant of the quadratic equation $p(y)=0$ becomes negative for $0<\beta_1<-\beta_2^2/\beta'_1$, allowing for the existence of two complex conjugate zeros.}).

\smallskip

Therefore, the $\zeta$-function for the massless case is regular at the origin and presents simple poles at  half-integer values of its argument: at $s=1/2$ with residue $\mu l/2\pi$ (linear in $l$), and at $s=\frac{1-2m}{2}, \, m=1,2,3,\cdots$, with residues $\sum_k \frac{(-1)^m \mathfrak{z}_k^{2 m-1}}{2 \pi  (2 m-1)}$ (independent of $l$). Thus, in this case $\zeta_A(s)$ also shows a singularity structure which follows the general rule valid for a second order differential operator subject to local boundary conditions \cite{Gilkey}.

\section{The determinant of $A$ for the massless case}

Since $\zeta_A(s)$ is analytic at the origin, the determinant can be defined as $\log {\rm Det} A:=-\zeta_A'(0)$. From Eqs.~(\ref{m-6}), (\ref{m-8}) and (\ref{m-16}) we have
\bea\label{m-17}
    &&\zeta_A(s)=-\mu l s +
    \sum_k \left\{ \frac{1}{2} -  s\,  \log \left(1-\mathfrak{z}_k \right)     \right\}\\
    &&+s \int_1^\infty dy\,  \frac{d}{dy}\log\left[1- e^{-2\mu l y } q(y)\right]
      + s \int_0^1 dy\, \frac{d}{dy}\log\left(\frac{g(i y)}{y}\right)\nonumber\\ &&- \frac{1}{2} + O\left( s^2 \right)\nonumber\\
      &&= -\frac{1}{2}\left[1- \sum_k 1 \right]+ s\left\{-\mu l  - \sum_k   \log \left(1-\mathfrak{z}_k \right)\right.\nonumber\\
      &&\left.-\log\left[1- e^{-2\mu l } \frac{p(-1)}{p(1)}\right]  + \log\left[\frac{ e^{\mu l} p(1)-e^{-\mu l} p(-1) }{2\mu(l \beta_1-\beta_2)}\right]
      \right\} + O\left( s^2 \right)\nonumber\\
      &&=-\frac{1}{2}\left[1- \sum_k 1 \right]+ s \log \left[ \frac{p(1)}{2\mu(l \beta_1-\beta_2) \prod_k \left(1-\mathfrak{z}_k \right) }\right]
       + O\left( s^2 \right),\nonumber
\eea
where the coefficient of the linear in $s$ term is the opposite of $\log {\rm Det} A$. From Eq.~(\ref{m-13}) we get
\begin{equation}\label{detm0}
 \log {\rm Det} A = -\zeta'(0)=\left\{
    \begin{array}{l}\displaystyle
    -\log \left( \frac{\beta'_2 \mu^2}{2(l \beta_1-\beta_2)}\right),
    \quad \beta'_2\neq 0,
         \\ \\ \displaystyle
    -\log\left( \frac{\beta'_1 \mu}{2(\beta_2 - l \beta_1)} \right),
    \quad \beta'_2=0, \beta'_1\neq 0.
    \end{array}
    \right.
\end{equation}
Notice that, even though  $\zeta_A(0)$ is independent of the external mass scale, the determinant of $A$ do depend on $\mu$.

\section{The Casimir energy for the massless case}

The Casimir energy for the massless case is defined from Eq.~(\ref{26}) in terms of the Laurent expansion around $s=-1/2$ of the meromorphic extension of the $\zeta$-function associated to $A$,  originally defined for $\frac{1}{2}<\Re (s)<1$ as (See the last lines of Eqs.~(\ref{m-6}) and (\ref{m-8}))
\bea\label{m-18}
    &&\zeta_A(s) =
       \frac{\sin (\pi s)}{\pi}\left\{  \int_1^\infty dy\, y^{-2s} \left[
        \mu l + \frac{p'(y)}{p(y)} + \frac{d}{dy}\log\left[1- e^{-2\mu l y } q(y)\right]
       \right] \right.\nonumber\\
        &&\left.  +
        \int_0^1 dy\, y^{-2s} \frac{d}{dy}\log\left(\frac{g(i y)}{y}\right) - \frac{1}{2s} \right\}.
\eea

Taking into account Eqs.~(\ref{m-8}), (\ref{m-12}) and (\ref{m-16}), we get
\bea\label{m-19}
    &&\zeta_A(s)   = \frac{\mu l}{2 \pi} +\frac{1}{\pi } \sum_k \left\{
         -\frac{\mathfrak{z}_k }{ \left(2s+1\right)} + {1+\mathfrak{z}_k  \log (1-\mathfrak{z}_k )}
         \right\}\\
        &&-   \frac{1}{\pi}
        \Bigg\{
        \int_1^\infty dy\, y\,  \frac{d}{dy}\log\left[1- e^{-2\mu l y } q(y)\right]\nonumber\\ 
        &&+
         \int_0^1 dy\, y\, \frac{d}{dy}\log\left(\frac{2 i g(i y)}{y}\right) +1  \Bigg\}+O \left(s+\frac{1}{2}\right)\nonumber\\
         &&= \frac{\mu l}{2 \pi} +\frac{1}{\pi } \sum_k \left\{
         -\frac{\mathfrak{z}_k }{ \left(2s+1\right)} + {1+\mathfrak{z}_k  \log (1-\mathfrak{z}_k )}
         \right\}    -   \frac{1}{\pi}\nonumber\\
           &&-   \frac{1}{\pi}
        \left\{- \log\left[1- e^{-2\mu l } q(1)\right]  - \int_1^\infty dy\, \log\left[1- e^{-2\mu l y } q(y)\right]
        \right\}\nonumber\\
           &&-   \frac{1}{\pi}
        \left\{  \mu l+ \log p(1)+ \log\left(1-e^{-2\mu l} q(1)\right)
       - \int_0^1 dy\, \log\left(\frac{2 i g(i y)}{y}\right)
        \right\}\nonumber\\
        &&+ O \left(s+\frac{1}{2}\right)\nonumber\\
        &&= -\frac{\left(\sum_k \mathfrak{z}_k \right)}{\pi(2s+1)} -\frac{\mu l}{2 \pi} +
        \frac{1}{\pi } \sum_k \left\{ 1+{\mathfrak{z}_k  \log (1-\mathfrak{z}_k )}\right\} - \frac{\log p(1)}{\pi}\nonumber\\
        &&+\frac{1}{\pi}  \int_1^\infty dy\, \log\left[1- e^{-2\mu l y } q(y)\right]
        +\frac{1}{\pi}    \int_0^1 dy\, \log\left[{e^{\mu l y}p(y)-e^{-\mu l y}p(-y)}\right] \nonumber\\
        &&= -\frac{\left(\sum_k \mathfrak{z}_k \right)}{\pi(2s+1)} +
        \frac{1}{\pi } \sum_k \left\{ 1+{\mathfrak{z}_k  \log (1-\mathfrak{z}_k )}\right\} - \frac{\log p(1)}{\pi}\nonumber\\
        &&+
        \frac{1}{\pi}    \int_0^1 dy\, \log\left[{p(y)}\right]+\frac{1}{\pi}  \int_1^\infty dy\, \log\left[1- e^{-2\mu l y } q(y)\right]\nonumber\\
        &&+\frac{1}{\pi}    \int_0^1 dy\, \log\left[{1-e^{-2\mu l y}q(y)}\right],\nonumber
\eea
where we have integrated by parts the integrals in the second line. 

Note that the (singular and finite) terms in the first line of the right hand side of Eq.~(\ref{m-19}) are independent of $l$. So, its contribution can be removed by means of a renormalization of the zero energy level (remember that $p(y)$ is $l$-independent). Moreover, from Eqs.~(\ref{cotas2}) and (\ref{cotas10}), it follows that the first integral in the second line is $O\left(e^{-2\mu l }\right)$ for large $l$, while the second one is $O\left({1}/{\mu l}\right)$.

Therefore, if we define the Casimir energy by subtracting the energy corresponding to the \emph{free} ($l \rightarrow \infty$) space, we finally get
\begin{equation}\label{m-20}
    \begin{array}{c}\displaystyle
      E_{Cas}(l)= \frac{\hbar \mu}{2\pi} \int_0^\infty dy\, \log\left[1- e^{-2\mu l y } q(y)\right],
      \\ \\ \displaystyle
      =\frac{\hbar}{2\pi l} \int_0^\infty dt\, \log\left[1- e^{-2 t } q(t/\mu l)\right],
    \end{array}
\end{equation}
where we have assumed that $p(y)$ has no positive zeroes. Note that $ E_{Cas}(l) \rightarrow 0$ as $1/l$ for large values of $l$, and that the expression in the second line shows that $ E_{Cas}(l)$  is independent of the external mass scale introduced by the regularization, since $q(t/\mu l)$ does not depend on $\mu$ (See Eq.~(\ref{m-9})). 

\begin{figure}
    \begin{tikzpicture}
    \node(a){\includegraphics[width=0.8\textwidth]{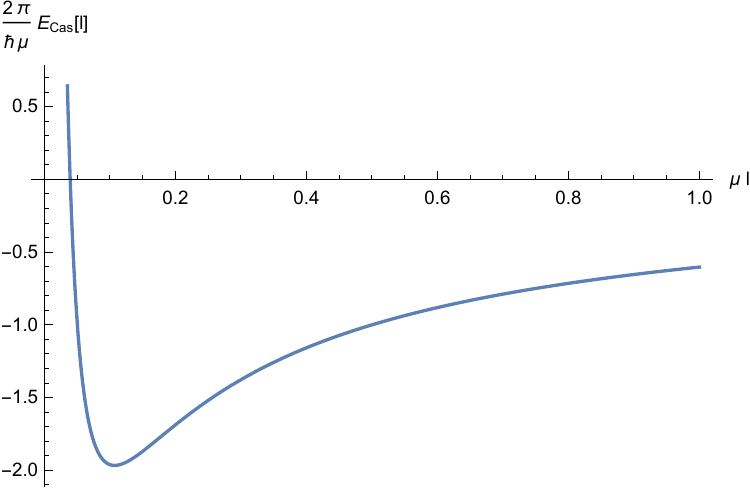}};
    \node at (a.north east)
    [
    anchor=center,
    xshift=-35mm,
    yshift=-10mm
    ]
    {
        \includegraphics[width=0.45\textwidth]{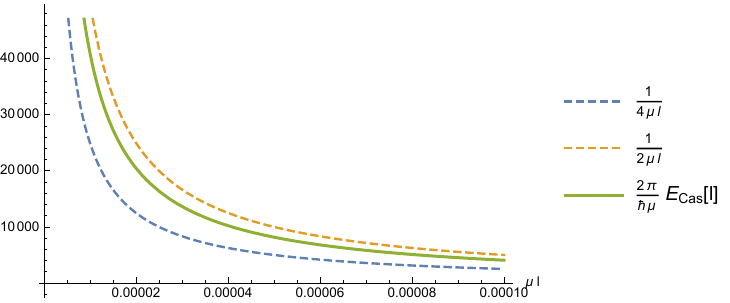}
    };
    \end{tikzpicture}
    \caption{$E_{Cas}$ as a function of $\mu l$ for $ b_1=7, b_2=-3, b'_1=-7, b'_2=1$.}
     \label{Ecas3m0}
\end{figure}

\begin{figure}
    \begin{tikzpicture}
    \node(a){\includegraphics[width=0.8\textwidth]{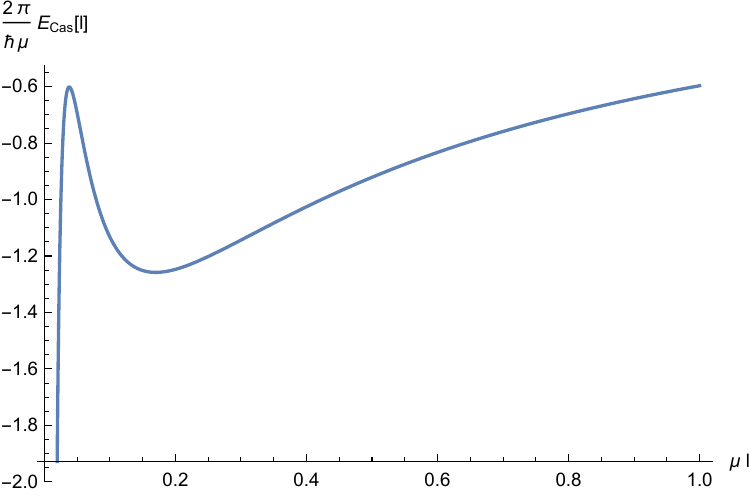}};
    \node at (a.north east)
    [
    anchor=center,
    xshift=-35mm,
    yshift=-55mm
    ]
    {
        \includegraphics[width=0.45\textwidth]{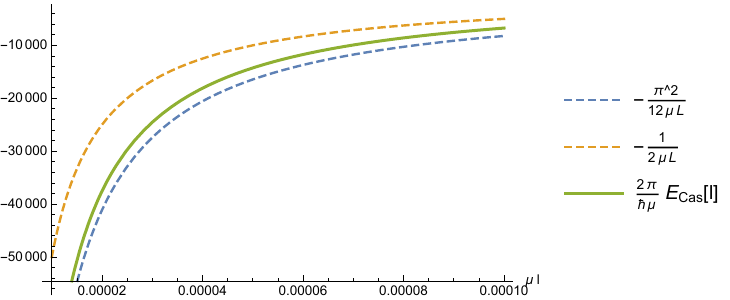}
    };
    \end{tikzpicture}
    \caption{$E_{Cas}$ as a function of $\mu l$ for $ b_1=81, b_2=-30, b'_1=-1, b'_2=0$.}
     \label{Ecas4m0}
\end{figure}

In Appendix~(\ref{bounds_m0}) we analyze the behavior of $E_{Cas}(l)$ for small values of $\mu l$, finding bounds similar to those found for the massive case: the Casimir force is strongly attractive or repulsive according with $\beta'_2=0$ or $\beta'_2\neq 0$. We illustrate this in Figures~(\ref{Ecas3m0}) and (\ref{Ecas4m0}) for different choices of the boundary parameters. Here again, local extrema in $E_{Cas}(l)$ are due to the existence of zeroes of $p(-y)$ in $(0,\infty)$. In these figures, $E_{Cas}$ has been plotted as a function of $\mu l$, and the parameters $\beta$ have been scaled with the power of $\mu$ corresponding to their dimension: $\beta_1=b_1 \mu^2,  \beta_2=b_2 \mu,  \beta'_1=b'_1 \mu^0, \beta'_2=b'_2/\mu$ with $b_i$ pure numbers.

\medskip

\section{Conclusions}\label{Conclusions}

In this article, we have studied the dynamics of a scalar field confined to a segment $[0, l]$, subject to dynamical boundary conditions at one end. These conditions depend not only on the local values of the field and its first spatial (normal) derivative but also on its first- and second-order time (tangential) derivatives. The search for stationary solutions reduces the dynamical equation to a boundary eigenvalue problem, where a modified Sturm-Liouville second-order differential operator $A$ is defined on a Hilbert space generated by eigenfunctions that satisfy boundary conditions dependent on the spectral parameter $\omega$.

We have analyzed the meromorphic structure of $\zeta_A(s)$, the $\zeta$-function associated with this operator, as a function of the parameters appearing in the boundary condition at $z = l$ (Eq.~(\ref{9}), assuming Dirichlet boundary conditions at $z = 0$). The singularities of $\zeta_A(s)$ in the complex plane are determined by the asymptotic behavior of the eigenvalues. We considered both the massive and massless cases and found that $\zeta_A(s)$ exhibits isolated simple poles located at $s = (1-n)/2 \notin \mathbb{Z}_-$, where $n = 0, 1, 2, \cdots$. This result aligns with the general rule for second-order differential operators subject to standard local boundary conditions.

This behavior has been explained by noting that, for large eigenvalues, the contributions from the time-derivative terms in the boundary conditions dominate. Consequently, they effectively impose standard local boundary conditions at $z = l$: generalized Robin boundary conditions when $\beta'_2 \neq 0$, and Dirichlet boundary conditions when $\beta'_2 = 0$.

In both cases, $\zeta_A(s)$ is holomorphic in a neighborhood of the origin, which allows the functional determinant of $A$ to be defined as $e^{-\zeta'_A(0)}$.

The analytic continuation of $\zeta_A(s)$ around $s = -1/2$ reveals that, in the massive case, the Casimir energy requires the renormalization of both a singular constant term and a singular term linear in $l$. The former corresponds to a redefinition of the reference energy level, while the latter involves subtracting a divergent energy density.

Since this procedure is performed with reference to a given external mass scale $\mu$, it leaves behind undetermined coefficients in a constant and a linear term in $E_{Cas}(l)$, a contribution of the form $\mathcal{E}_0+\mathcal{E}_1 l $. In particular, both $\mathcal{E}_0$ and $\mathcal{E}_1$ run with respect to $\mu$ to ensure the independence of $E_{Cas}(l)$ with respect to this external mass scale introduced by the analytic regularization.

The finite contribution to $E_{Cas}(l)$ is expressed as an integral suitable for numerical evaluation, which is observed to decay exponentially fast for large $l$. Consequently, the linear term in $E_{Cas}(l)$ represents an energy density that, if nonzero, governs the expansion or contraction of the manifold (segment) for large values of $l$. In this sense, $\mathcal{E}_1$ resembles a renormalized cosmological constant.

On the other hand, since the exact analytic behavior of $E_{Cas}(l)$ for small values of $l$ could not be determined, we provide estimates for this limit. Our analysis indicates that, for $\beta'_2 \neq 0$, $E_{Cas}(l)$ is strongly repulsive when $m l \ll 1$, while for $\beta'_2 = 0$, it is strongly attractive. This behavior resembles the Casimir energy for standard generalized Robin (including Neumann) and Dirichlet boundary conditions, respectively.  Furthermore, we include figures to illustrate these behaviors, which show that $E_{Cas}(l)$ can also exhibit local extrema at finite values of $l$, depending on the choice of boundary parameters.

For the massless case, the analytic continuation of $\zeta_A(s)$ around $s = -1/2$ requires only the subtraction of a singular constant term. The finite contribution is expressed as an integral suitable for numerical evaluation, which behaves as $E_{Cas}(l) = O(1/l)$ for large values of $l$ and is manifestly independent of $\mu$.

In this case, we also derive estimates for the small $l$ limit, obtaining results similar to those for standard local boundary conditions, depending on whether $\beta'_2$ vanishes or not, as previously described. To illustrate these behaviors of $E_{Cas}(l)$, we include figures that also demonstrate the possibility of local extrema depending on the choice of boundary parameters.


\section*{Acknowledgments}
H.F.\ thanks the kind hospitality of Universidad San Sebastián and Pontificia Universidad Católica de Santiago, Chile. H.F.\ also thanks support from UNLP under Grant 11/X909 and from CONICET under grant PIP KE-11220220100262CO, Argentina. 
E. M.\ thanks support from Fondecyt No 1230440, Chile. M. L thanks support from Fondecyt No. 1220035 and No 1241436, Chile.

\appendix

\section{Iterative evaluation of $\zeta_A(s)$ for the $\beta'_2=0$ case}\label{betap20}
Let us consider the iterative construction of eigenvalues for the $\beta'_2=0$ ($b=0$) case. From Eq.~(\ref{12-2}) we get
\begin{equation}\label{betap20-1}
        \frac{ \tan x  }{x}= \frac{a}{c+d \, x^2 } \approx \left\{
    \begin{array}{l}
      \frac{a}{c}  +O(x^2), \quad {\rm for} \quad x<<1, \, c \neq 0,
      \\ \\
       \frac{a}{d x^2}>0, \quad {\rm for }\quad x \rightarrow \infty,\, \beta'_1 \neq 0.
    \end{array}
    \right.
\end{equation}
Therefore, for large $n$ we posit an asymptotic expansion in the form 
\begin{equation}\label{betap20-2}
  x_n \asymp n \pi + \sum_{k=1}^K \frac{y_k}{n^k} +O\left(n^{-K-1}\right),
\end{equation}
and obtain
\begin{equation}\label{betap20-3}
    \begin{array}{l}\displaystyle
         y_1=\frac{\beta_2 l}{\pi \beta'_1},        \quad y_2=0,
         \\ \\ \displaystyle
         y_3=-\frac{\beta_2 l^2 } {3 \pi ^3 {\beta'_1}^3} 
         \left\{3 \beta_2 \beta'_1 +l \left[{\beta_2}^2+3 \beta'_1
   \left(\beta_1 +\beta'_1 m^2\right)\right]\right\} , 
   \\ \\ \displaystyle
   y_4=0, \quad \cdots
    \end{array}
\end{equation}
This leads to the approximation to the $\zeta$-function
\begin{equation}\label{betap20-4}
    \zeta_A(s)\approx \left(\frac{\mu l}{\pi}\right)^{(2s)}
    \left\{ \zeta(2s)-  \frac{ sl}{\pi^2 \beta'_1} \left( 2 \beta_2+ l m^2 \beta'_1\right)\zeta(2s+2) 
    \right\} + \Delta\zeta(s),
\end{equation}
from which we get for the determinant of $A$
\bea\label{betap20-5}
     &&\log{\rm Det} (A)={-{\zeta_A}'(0)}=-\ln\left(2\mu l\right) + 
     \frac{l}{540 {\beta'_1}^3} \left\{
    -180 \beta_2  {\beta'_1}^2\right.\nonumber\\
   &&\left. +3
  {\beta'_1}^3 l^3 m^4 +4 \beta_2  l^2
   \left[{\beta_2}^2+3\beta'_1
   \left(\beta_1+2\beta'_1
   m^2\right)\right]\right.\nonumber\\
   &&\left.+18\beta'_1 l
   \left({\beta_2}^2-5{\beta'_1}^2
   m^2\right) \right\} + \Delta\zeta'(0),
\eea
and the contribution to the Casimir energy
\bea
    &&\left.\frac{\hbar \mu}{2} \zeta_A(s)\right|_{s\approx -\frac 1 2}
    = \frac{\hbar  \left(2 \beta_2 +\beta'_1 l m^2\right)}{2 \pi  \beta'_1  \left(s+\frac{1}{2}\right)} 
    -\frac{\pi  \hbar }{24 l}
    +\frac{ \hbar \beta_2  \log \left(\frac{\mu 
   L}{\pi }\right)}{2 \pi  \beta'_1}\nonumber\\
   &&+O\left((\mu l)^0\right)+O\left(s+1/2\right).
\eea

\section{Bounds of $E_{Cas}(l)$ for small $l$ with $m=0$}\label{bounds_m0} 

In this appendix we establish bounds for the Casimir energy for small $l$ in the massless case, which we write as  
\begin{equation}\label{m0-1}
   \frac{2\pi}{\hbar \mu}  E_{Cas}(l)=\int_0^{\infty} dy\, \ln[1-e^{-2\mu l y} q(y)] = I_1(l)+I_2(l)
\end{equation}
where
\begin{equation}\label{m0-2}
    \begin{array}{l}\displaystyle
        I_1(L):= \int_0^{1} dy\, \ln[1-e^{-2\mu l y} q(y)] ,
        \\ \\  \displaystyle
        I_2(l):= \int_1^{\infty} dy\, \ln[1-e^{-2\mu l y} q(y)] ,
        \\ \\  \displaystyle
        q(y):=\frac{\beta_1+\beta_2 \mu  y-\beta'_1\mu^2 y^2-\beta'_2 \mu^3 y^3}{\beta_1-\beta_2 \mu  y-\beta'_1\mu^2 y^2+\beta'_2 \mu^3 y^3}.
    \end{array}
\end{equation}
In so doing we must take into account not only the behavior of $q(y)$ for large values of $y$,
\begin{equation}\label{m0-3}
    \lim_{y\rightarrow\infty} q(y)=\left\{
    \begin{array}{ll}
        -1, & {\rm for}\ \beta'_2 \neq 0 ,\\ \\
         +1,  &  {\rm for}\ \beta'_2 =0,
    \end{array}\right.
\end{equation}
but also its behavior near the origin,
\begin{equation}\label{m0-4}
    \lim_{y\rightarrow 0} q(y)=\left\{
    \begin{array}{ll}
        1, & {\rm for}\ \beta_1 \neq 0 ,\\ \\
        -1,  &  {\rm for}\ \beta_1 =0,
    \end{array}\right.
\end{equation}

We consider first $I_2(l)$ and then $I_1(l)$. We show that the singular behavior of $I_2(2)$ for small $l$ is dominant and similar to the one found for the massive case, while the corresponding behavior of $I_1(l)$ is subdominant.

\subsection{The contribution of $I_2(l)$ for $\mu l\ll 1$}

\subsubsection{ $\lim_{y \rightarrow\infty}{q(y)}=-1$ ($\beta'_2\neq 0$)}

Here we consider the case $\lim_{y \rightarrow\infty}{q(y)}=-1$. Let $y_0>1$ be sufficiently large so that $-1-\epsilon <q(y)<-1+\epsilon$ for all $y>y_0$. Notice that $q(y)$ is independent of $l$, and so is $y_0$. Indeed, for large $y$ in this case we have
\begin{equation}\label{q-large-y}
    q(y)+1= -\frac{2\beta'_1}{\beta'_2 \mu y} +O\left(y^{-2}\right).
\end{equation}

Then, we have
\bea\label{m0-5}
    \int_{y_0}^\infty dy\, \ln[1+ (1-\epsilon)e^{-2\mu l y}]&<& 
     \int_{y_0}^\infty dy\,  \ln[1- q(y)e^{-2\mu l y} ]\nonumber\\
     &<& 
      \int_{y_0}^\infty dy\, \ln[1+(1+\epsilon)e^{-2\mu l y} ],
\eea
since the logarithm is a monotonously increasing function. From the inequalities in Eq.\ (5.11) we can write for the integral in the right hand side
 \bea\label{m0-6}
     \int_{y_0}^\infty dy\,  \ln[1+ (1+\epsilon) e^{-2\mu l y} ] &<&
       (1+\epsilon) \int_{y_0}^\infty dy\,  e^{-2\mu l y}\nonumber\\ &=&
        \frac{(1+\epsilon)}{2\mu l} e^{-2\mu l y_0} 
 \eea
 $\forall \epsilon>0$, in such a way that  
 \begin{equation}\label{m0-7}
     \int_{y_0}^\infty dy\,  \ln[1- q(y)e^{-2\mu l y} ] 
< \frac{e^{-2\mu l}}{2\mu l}  =\frac{1}{2\mu l} + O\left((\mu l)^0\right).
 \end{equation}
 On the other hand, 
 \begin{equation}\label{m0-8}
 \begin{array}l \displaystyle
 \int_{y_0}^\infty dy\, \ln[1+ (1-\epsilon)e^{-2\mu l y}]>
     \int_{y_0}^\infty dy\, \frac{(1-\epsilon)e^{-2\mu l y}}{1+ (1-\epsilon)e^{-2\mu l y}}
     \\ \\ \displaystyle
     =\frac{1}{2\mu l} \ln[1+(1-\epsilon)e^{-2\mu l y_0}]
      > \frac{(1-\epsilon)}{2\mu l}\, \frac{e^{-2\mu l y_0}}{1+e^{-2\mu l y_0}}
      \\ \\ \displaystyle
          > \frac{(1-\epsilon)}{4\mu l}\, e^{-2\mu l y_0} 
     = \frac{(1-\epsilon)}{4\mu l} +O\left((\mu l)^0\right),
 \end{array}
 \end{equation}
 $\forall \epsilon>0$ (small enough). 

 Then, we have
 \begin{equation}\label{m0-9}
      \frac{1}{4\mu l} + O\left((\mu l)^0\right) \leq 
     \int_{1}^\infty dy\,  \ln[1- q(y)e^{-2\mu l y} ]\leq 
     \frac{1}{2\mu l} + O\left((\mu l)^0\right),
\end{equation}
since $ \int_{1}^{y_0} dy\,  \ln[1- q(y)e^{-2\mu l y} ]= O\left((\mu l)^0\right)$.

\subsubsection{$\lim_{y \rightarrow\infty}{q(y)}=1$ ($\beta'_2=0$)}
Now, we take $y_0$ sufficiently large so as $1-\epsilon <q(y)<1+\epsilon$ for all $y>y_0$. Then, 
\bea\label{m0-10}
    \int_{y_0}^\infty dy\, \ln[1- (1+\epsilon)e^{-2\mu l y}]&<& 
     \int_{y_0}^\infty dy\,  \ln[1- q(y)e^{-2\mu l y} ]\nonumber\\
     &<& 
      \int_{y_0}^\infty dy\, \ln[1-(1-\epsilon)e^{-2\mu l y} ].
\eea
 From Eq.\ (5.19), for the right hand side we can write 
 \begin{equation}\label{m0-12}
 \begin{array}{l}\displaystyle
     \int_{y_0}^\infty dy\,  \ln[1- (1-\epsilon) e^{-2\mu l y} ] <
      - (1-\epsilon) \int_{y_0}^\infty dy\,  e^{-2\mu l y} 
      \\ \\ \displaystyle
    =- \frac{(1-\epsilon)}{2\mu l} e^{-2\mu l y_0} \leq 
    - \frac{e^{-2\mu l y_0}}{2\mu l} = \frac{-1}{2\mu l} +O\left((\mu l)^0\right) .
 \end{array}
 \end{equation}
 On the other hand, for the left hand side we have
 \bea\label{m0-13}
 &&\int_{y_0}^\infty dy\, \ln[1- (1+\epsilon)e^{-2\mu l y}]=
     \int_{y_0}^\infty dy\, \ln[1- e^{-2\mu l y}]\nonumber\\
     &&+
     \int_{y_0}^\infty dy\, \ln\left[1- \frac{\epsilon }{e^{2\mu l y}-1 }\right]\nonumber\\
      &&> \int_{y_0}^\infty dy\, \ln[1- e^{-2\mu l y}]- \epsilon
     \int_{y_0}^\infty dy\, \frac{1}{e^{2\mu l y}-2}\nonumber\\  
     &&= -\frac{\pi ^2}{12 \mu  l}+(1-\log (2 \mu l))+O\left((\mu l)^1\right)
     -\epsilon  \int_{y_0}^\infty dy\, \frac{1}{e^{2\mu l y}-2}
 \eea
 $\forall \epsilon>0$ (sufficiently small). Since the last integral takes positive values, and
 \begin{equation}\label{m0-14}
     \int_1^{y_0} dy\,  \ln[1- q(y)e^{-2\mu l y} ] =
      \int_1^{y_0} dy\,  \ln[1- q(y) ] +O\left((\mu l)^1\right)
 \end{equation}
we can write 
 \bea\label{m0-15}
     -\frac{\pi ^2}{12 \mu  l}-\log (\mu l)+O\left((\mu l)^0\right) &\leq&
      \int_{y_0}^\infty dy\,  \ln[1- q(y)e^{-2\mu l y} ]\nonumber\\
      &\leq&      - \frac{1}{2\mu l} +O\left((\mu l)^0\right),
 \eea

Notice that these asymptotic behaviors for small distances are similar to those found for the massive case.

\subsection{The contribution of $I_1(l)$ for $\mu l \ll 1$} 
We now consider the contribution of from $I_1(l)$ to the Casimir energy  for small values of $\mu l$. 

\subsubsection{ $\lim_{y\rightarrow 0}q(y)=1$ ($\beta_1\neq 0$)}
Taking into account that the derivative of $q(y)$ at the origin is 
\begin{equation}\label{m0-16}
    q'(0)=2\mu \beta_2/\beta_1 <0
\end{equation}
(See Eq.\ (1.10)), we can consider $\epsilon>0$  and $y_1$ sufficiently samll so as $1-\epsilon < q(y) \leq 1$ for all $0\leq y<y_1<1$. Then, 
\bea\label{m0-17}
    \int_0^{y_1}dy\, \ln(1-e^{-2\mu l y})&\leq& \int_0^{y_1}dy\, \ln(1-e^{-2\mu l y}q(y))\nonumber\\
    &<&
    \int_0^{y_1}dy\, \ln(1-(1-\epsilon)e^{-2\mu l y}).
\eea
For the left hand side 
\begin{equation}\label{m0-18}
\begin{array}{l}\displaystyle
     \int_0^{y_1}dy\, \ln(1-e^{-2\mu l y})= \int_0^{y_1}dy \left(\ln(2\mu l y) + O(\mu l)\right)
     \\ \\ \displaystyle
    =  y_1\left(\ln(2\mu l y_1)-y_1\right)+ O(\mu l)>\ln(\mu l) +  O\left((\mu l)^0\right),
\end{array}  
\end{equation}
since $\ln(\mu l)<0$ in the region of interest. On the other hand,
\begin{equation}\label{m0-19}
    \begin{array}{l}\displaystyle
        \int_0^{y_1}dy\, \ln\left(1-(1-\epsilon)e^{-2\mu l y}\right)< 
         -(1-\epsilon) \int_0^{y_1}dy\, e^{-2\mu l y} 
        \\ \\ \displaystyle
       =-\frac{(1-\epsilon)}{2\mu l}\left[1-e^{-2\mu l y_1}\right]=
O\left((\mu l)^0\right),
    \end{array}
\end{equation}
Therefore,
\begin{equation}\label{m0-20}
    \ln(\mu l) +  O((\mu l)^0)< \int_0^{1}dy\, \ln(1-e^{-2\mu l y}q(y))< O((\mu l)^0),
\end{equation}
since $\int_{y_1}^{1}dy\, \ln(1-e^{-2\mu l y}q(y))=O((\mu l)^0)$.

\subsubsection{ $\lim_{y\rightarrow 0}q(y)=-1$ ($\beta_1= 0$)}
In this case
\begin{equation}\label{m0-21}
    q'(0)=2\mu \beta'_1/\beta_2>0
\end{equation}
(See Eq.\ (1.10)). Then, given $\epsilon>0$, we can choose $0<y_1<1$ such that $-1\leq q(y)<-(1-\epsilon)$ for all $y\in[0,y_1)$, so as to write
\bea\label{m0-22}
    \int_0^{y_1}dy\, \ln\left(1+(1-\epsilon)e^{-2\mu l y}\right)&<&
    \int_0^{y_1}dy\, \ln\left(1-q(y)e^{-2\mu l y}\right)\nonumber\\
    &<&
    \int_0^{y_1}dy\, \ln\left(1+e^{-2\mu l y}\right)
\eea
from which we get
\bea\label{m0-23}
    y_1 \ln(2-\epsilon) + O(\mu l)&<&
    \int_0^{y_1}dy\, \ln\left(1-q(y)e^{-2\mu l y}\right)\nonumber\\
    &<&
    y_1 \ln(2)+ O(\mu l)<\ln(2) + O(\mu l).
\eea
 On the other hand,
 \begin{equation}\label{m0-24}
     \int_{y_1}^1 dy\, \ln\left(1-q(y)e^{-2\mu l y}\right)=
     \int_{y_1}^1 dy\, \ln\left(1-q(y))\right) + O(\mu l).
 \end{equation}
So,
 \begin{equation}\label{m0-25}
   \int_0^{1}dy\, \ln\left(1-q(y)e^{-2\mu l y}\right)= O\left((\mu l)^0\right).
\end{equation}

\bigskip
These bounds for $I_1(l)$ should be combined with those previously found for $I_2(l)$ for each choice of the $\beta'$s parameters, although in each case the later ones give the dominant contribution to the Casimir energy at small $\mu l$.


\section{$E_{Cas}(l)$ for the Sturm-Liouville problem with standard local boundary conditions} \label{local}

\begin{figure}
    \begin{tikzpicture}
    \node(a){\includegraphics[width=0.8\textwidth]{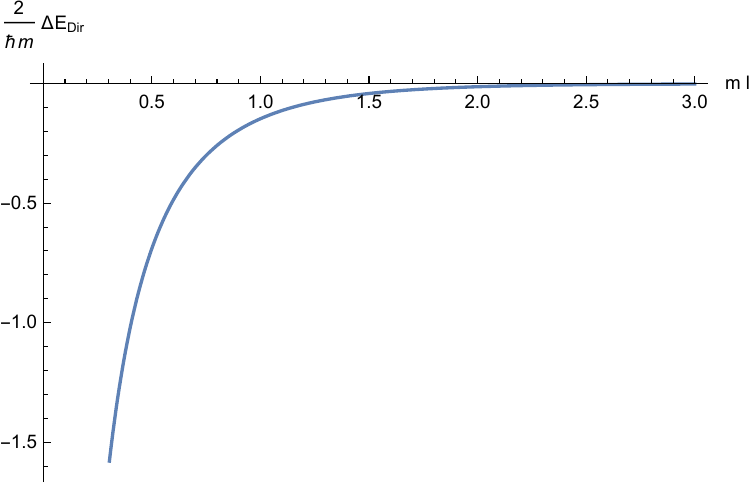}};
    \node at (a.north east)
    [
    anchor=center,
    xshift=-35mm,
    yshift=-45mm
    ]
    {
        \includegraphics[width=0.45\textwidth]{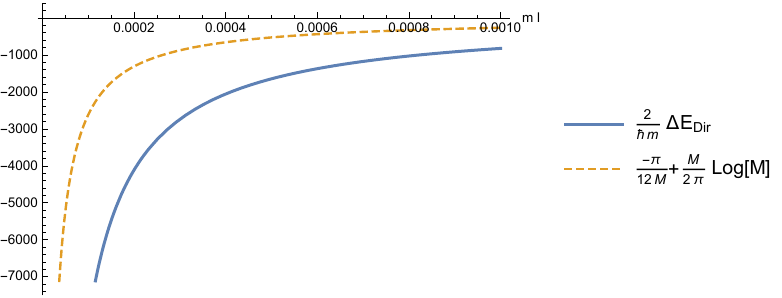}
    };
    \end{tikzpicture}
    \caption{$\Delta E_{Dir}$ for local boundary conditions as a function of $M= m l$, for $\beta_1=m^2,\beta_2=0$.}
     \label{Dirichlet-SL}
\end{figure}

For completeness, let us consider in this Appendix the Sturm-Liouville operator $A$ in Eq.\ (2.1) subject to the usual local boundary conditions at $z=l$ ({\it i.e.} with $\beta'_1=0=\beta'_2$). For Dirichlet boundary conditions ($\beta_2=0$) we have for the eigenvalues
\begin{equation}\label{local-1}
    \sin{\left(l\sqrt{\omega^2-m^2}\right)}=0 \  \rightarrow \  
    \omega_n=\sqrt{\left(\frac{n \pi}{l}\right)^2+m^2}, \ n=1,2,3 \dots,
\end{equation}
and for $\zeta(s)$
\begin{equation}\label{local-2}
\begin{array}{l}\displaystyle
     \zeta_D(s)=\left(\frac{\mu l}{\pi}\right)^{2s} 
    \sum_{n=1}^\infty n^{-2s}\left[1+\left(\frac{m l}{\pi n}\right)^2\right]^{-s}=
    \\ \\  \displaystyle
     = \left(\frac{\mu l}{\pi}\right)^{2s} 
    \sum_{n=1}^\infty n^{-2s}\left[1-\frac{l^2 m^2 s}{\pi ^2 n^2}
    +\frac{l^4 m^4 s (s+1)}{2 \pi ^4 n^4} +O(n^{-6})\right]=
     \\ \\  \displaystyle
     =\left(\frac{\mu l}{\pi}\right)^{2s} \left\{\zeta (2 s)
     -\frac{l^2 m^2 s}{\pi ^2} \zeta (2 (s+1)) + O(m l)^4 \right\}.
\end{array}
\end{equation}
Therefore, for $\Delta E$ in Eq.~(\ref{41}) we get in this case (See Figure~(\ref{Dirichlet-SL}))
\begin{equation}\label{local-3}
    E_{Dir}(l)= \frac{\hbar m^2 l}{8\pi(s+1/2)}
    -\frac{\hbar \pi}{24 l}+\frac{\hbar m^2}{4\pi} l \ln(\mu l) +O(s+1/2)
\end{equation}
which, in the limit $m\rightarrow 0$ reduces to the exact result for the massless case,
\begin{equation}\label{local-4}
    E_{Dir}(l)=     -\frac{\hbar \pi}{24 l}.
\end{equation}
Indeed, in this case the $\zeta$-function reduces to the series in the first line of Eq.~(\ref{local-2}) but involving only the first term inside the brackets. Notice that the dominant term in the small $l$ limit is strongly attractive. 

\bigskip

On the other hand, if we adopt general Robin boundary conditions at $z=l$ ($\beta_1,\beta_2\neq 0$), the eigenvalues satisfy
\begin{equation}\label{local-5}
\beta_2 \sqrt{\omega^2-m^2}\, \cos{\left(l\sqrt{\omega^2-m^2}\right)}-
    \beta_1 \sin{\left(l\sqrt{\omega^2-m^2}\right)}=0.
\end{equation}
An asymptotic expansion of the eigenvalues for $n\gg 1$ leads to the following approximation to the $\zeta$-function
\bea\label{local-6}
    &&\zeta_R(s)=\left(\frac{\mu l}{\pi}\right)^{2s} 
    \sum_{n=1}^\infty  n^{-2s}\left[
    1+\frac{s}{n} +\right.
    \nonumber\\
    &&\left. + \frac{s(s+1)}{12\pi^2 \beta_2 n^3}\left(
    24 l \beta_1+\beta_2[\pi^2(1+2s)-12 m^2 l^2]\right) +O(n^{-5})
    \right]\nonumber\\
     &&=\left(\frac{\mu l}{\pi}\right)^{2s} \left\{
     \zeta (2s)+s \zeta (2 s+1)\right.\nonumber\\ &&\left.+\frac{s}{4}\left(1+2s-\frac{4 l \left(\beta_2 l m^2-2    \beta_1\right)}{\pi ^2 \beta_2}\right) \zeta(2s+2)   
     \right\}.
\eea

\begin{figure}
\includegraphics[width=\linewidth]{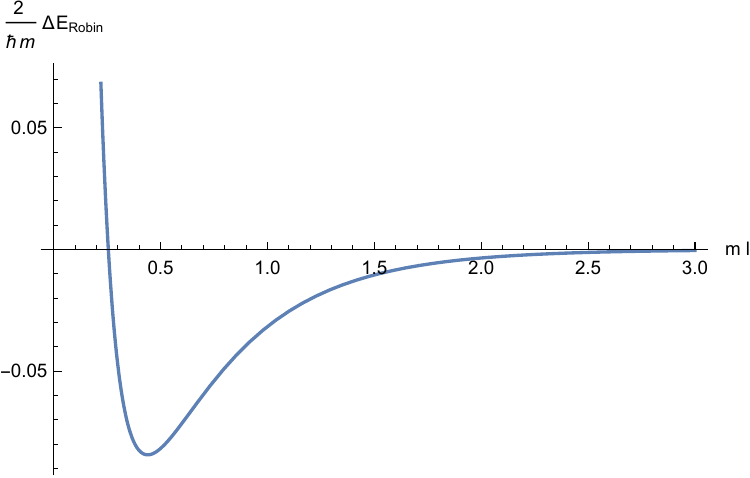}
\caption{$\Delta E_{Robin}$ for local boundary conditions as a function of $M=m l$, for $\beta_1=2 m^2,\beta_2=-m$.}
 \label{Robin-SL}
\end{figure}

\begin{figure}
    \begin{tikzpicture}
    \node(a){\includegraphics[width=0.8\textwidth]{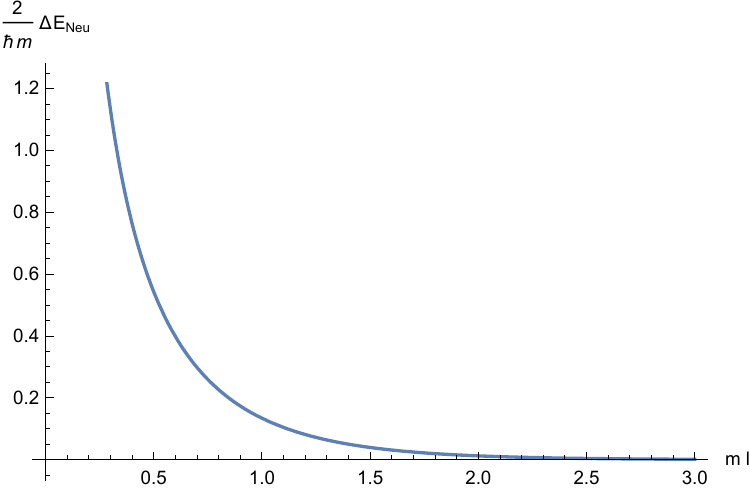}};
    \node at (a.north east)
    [
    anchor=center,
    xshift=-35mm,
    yshift=-30mm
    ]
    {
        \includegraphics[width=0.45\textwidth]{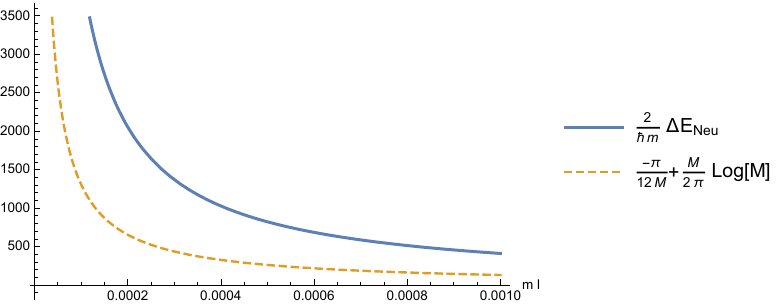}
    };
    \end{tikzpicture}
    \caption{$\Delta E_{Neu}$ for local boundary conditions as a function of $M=m l$, for $\beta_1=0,\beta_2=m$.}
     \label{Neumann-SL}
\end{figure}

In this case $E_{Rob}(l)$  behaves as (See Figure~(\ref{Robin-SL}))
\bea\label{local-7}
     E_{Rob}(l)= \frac{ \hbar(\beta_2 l m^2-2 \beta_1)}{8 \pi \beta_2  
   \left(s+\frac{1}{2}\right)}+\frac{\hbar \pi}{48 l}
   -\frac{\hbar \beta_1 \log (l \mu )}{2 \pi \beta_2}
   +\frac{l m^2 \hbar  \log (l \mu )}{4 \pi }+\cdots
\eea
which, for $\beta_1=0$ (Neumann boundary conditions) reduces to (See Figure \ref{Neumann-SL})
\begin{equation}\label{local-8}
    E_{Neu}(l)= \frac{ \hbar  l m^2}{8 \pi   
   \left(s+\frac{1}{2}\right)}+\frac{\hbar \pi}{48 l}
   +\frac{l m^2 \hbar  \log (l \mu )}{4 \pi }+\cdots
\end{equation}
In this case, the Casimir energy is strongly repulsive at short distances.

\section{A toy model with dynamical boundary conditions}

As an additional motivation for considering the generalized dynamical boundary conditions introduced in \cite{JA_W-1}, let us consider a simple model of a scalar field defined on a segment $[0,l]$, described by the Lagrangian
\bea
L = \frac{1}{2}\int_0^l dz \left[
\epsilon(z) \left( \frac{\partial \varphi}{\partial t} \right)^2
- \left( \frac{\partial \varphi}{\partial z} \right)^2 - V(z)\varphi^2
\right] \equiv \int_0^l\mathcal{L}dz.
\eea
The variational principle leading to Euler-Lagrange equations, 
\bea
\frac{\partial}{\partial t}\left( \frac{\partial\mathcal{L}}{\partial(\partial\varphi/\partial t)} \right)
+ \frac{\partial}{\partial z}\left( \frac{\partial \mathcal{L}}{\partial(\partial\varphi/\partial z)} \right) - \frac{\partial\mathcal{L}}{\partial \varphi} = 0,
\eea
allows us to obtain the corresponding equations of motion, i.e.
\bea
\epsilon(z) \frac{\partial^2 \varphi}{{\partial t}^2} - \frac{\partial^2\varphi}{\partial z^2} + V(z)\varphi(t,z) = 0,
\label{eq_EOM}
\eea
which we complement by imposing local (time independent) boundary condition at both ends of this segment. 

Let us now assume that the media represents a small junction of thickness $\Delta \ll l$, centered at $0<z_0<l$, connecting two separate regions with different static dielectric permitivities. This condition is modeled by a discontinuous function
\bea
\epsilon(z) = \left\{\begin{array}{cc} 
\epsilon_1, & z < z_0 - \Delta\\
\epsilon_{\Delta}, & z_0 - \Delta < z < z_0 + \Delta\\
\epsilon_2, & z > z_0 + \Delta
\end{array}\right.
\eea
which can be expressed more conveniently by means of the Heaviside function $\Theta(z)$ as follows,
\bea
\epsilon(z) &=& \epsilon_1 \Theta(z_0 - \Delta - z) + \epsilon_2 \Theta(z - z_0 - \Delta)\nonumber\\
&+& \epsilon_{\Delta}\left[ \Theta(z - z_0 + \Delta) - \Theta(z - z_0 - \Delta) \right].
\label{eq_epsx}
\eea
In the limit of a very narrow junction, $\Delta/l \ll 1$, we can approximate the dielectric permittivity $\epsilon(z)$ through a {\it{distributional}} Taylor expansion as 
\bea
\epsilon(z) &\sim& \epsilon_1 + \left( \epsilon_2 - \epsilon_1 \right)\Theta(z-z_0)
+ \Delta\left( 2\epsilon_{\Delta} + \epsilon_1 - \epsilon_2 \right) \, \delta(z-z_0)\nonumber\\
&+& \frac{\Delta^2}{2}\left( \epsilon_2 - \epsilon_1  \right) \, \delta'(z - z_0) + \ldots
\eea
up to $O\left(\frac{\Delta}{l}\right)^2$,  where we used the parity properties $\delta(z-z_0) = \delta(z_0-z)$, and $\delta'(z-z_0) = -\delta'(z_0-z)$.
From now on, we shall also assume that the local potential develops a kink at the junction $z=z_0$, such that it can be represented by the distribution
\bea
V(z) \rightarrow m^2 + v(z) - V_0\delta(z-z_0).
\eea
Then, from \ref{eq_EOM} we get 
\bea
\left[\epsilon_1 + \left( \epsilon_2 - \epsilon_1 \right)\Theta(z-z_0)+
R \ \delta(z-z_0)+ S\ \delta'(z - z_0)\right]\partial_t^2{\varphi}(t,z)  \nonumber \\
- {\partial_z^2\varphi}(t,z)  + \left[ v(z) - V_0\delta(z-z_0) \right]\varphi(t,z) = 0,
\label{eq_EOM-model}
\eea
where $R= \Delta\left( 2\epsilon_{\Delta} + \epsilon_1 - \epsilon_2 \right)$ and $S=\frac{\Delta^2}{2}\left( \epsilon_2 - \epsilon_1  \right)$. Since the right hand side is regular (identically vanishing), the singular terms in the left hand side must cancel out. This requires that $\partial_z\varphi(t,z)$ be discontinuous at $z=z_0$, in order to cancel the $\delta$-term in the potential,
\be
\partial_z \varphi(t,z_0^+)-\partial_z \varphi(t,z_0^-)= V_0 \varphi(t,z_0),
\ee
where $\varphi(t,z_0)$ is a well defined value, since $\varphi(t,z)$ must be a continuous function for all $t$.  This introduces an indeterminacy in the $\delta'$-term, which we circumvent by defining (we take $\delta(x) \Theta[x]:=\frac12  \delta(x)$)
\be
\int_{z_0^-}^{z_0^+} \delta'(z - z_0) \partial_t^2{\varphi}(t,z) dz:= - \frac{1}{2}\partial_t^2 \left[ \partial_z\varphi(t,z_0^+)-\partial_z\varphi'(t,z_0^-) \right],
\ee

So, by integrating both sides of Eq.~(\ref{eq_EOM-model}) we get
\bea
&&\int_{z_0^-}^{z_0^+} \left\{
\left[\epsilon_1 + \left( \epsilon_2 - \epsilon_1 \right)\Theta(z-z_0)+
R \ \delta(z-z_0)+ S\ \delta'(z - z_0)\right]\partial_t^2{\varphi}(t,z) \right. \nonumber\\
&&\left. - \partial_z^2{\varphi}(t,z)  + \left[ v(z) - V_0\delta(z-z_0) \right]\varphi(t,z)\right\} dz = R\, \partial_t^2{\varphi}(t,z_0)  \\
&&-\frac{S}{2} \partial_t^2 \left[ {\partial_z\varphi}(t,z_0^+)-{\partial_z\varphi}(t,z_0^-) \right]
-\left[ \partial_z\varphi(t,z_0^+)-\partial_z\varphi(t,z_0^-) \right]\nonumber\\ 
&&-V_0 {\varphi}(t,z_0)
=0, \nonumber
\label{int-eq_EOM}
\eea
or
\bea \label{chanchada}
\partial_t^2\left[R {\varphi}(t,z_0) +\frac{S}{2} {\partial_z\varphi}(t,z_0^-) \right] + \partial_z\varphi(t,z_0^-)-V_0 {\varphi}(t,z_0)
\nonumber \\
= \left[1+\frac{S}{2}   \partial_t^2  \right] {\partial_z\varphi}(t,z_0^+).
\eea
Now, if we choose Neumann boundary conditions at $z=l$ (${\partial_z\varphi}(t,l)=0, \forall t$ ), and taking into account that $\partial_z\varphi(t,z)$ is continuous for $z_0<z<l$, in  the limit $z_0 \rightarrow l^-$ we conclude that the scalar field satisfy the effective boundary conditions 
\be \label{BC-JA-Weder}
\partial_t^2\left[R {\varphi}(t,z_0) +\frac{S}{2} {\partial_z\varphi}(t,z_0) \right] =V_0 {\varphi}(t,z_0)- \partial_z\varphi(t,z_0),
\ee
which can be mapped onto the dynamical boundary conditions considered in \cite{JA_W-1} by identifying the parameters $\beta'_1=R$, $\beta'_2=-S/2$, $\beta_1=V_0$ and $\beta_2=1$.

This toy model serves as a physical example where these general dynamical boundary conditions are explicitly realized. 


\pagebreak
\section*{References}


\end{document}